\documentclass[fleqn,usenatbib,useAMS]{mnras}

\usepackage{graphicx}	%
\usepackage{amsmath}	%
\usepackage{amssymb}	%
\usepackage{multicol}        %
\usepackage{bm}		%
\usepackage{pdflscape}	%

\usepackage[T1]{fontenc}
\usepackage{ae,aecompl}

\usepackage{newtxtext,newtxmath}

\title[Cosmic web: shock obliquity and cosmic rays]{Shock waves in the magnetized cosmic web: the role of obliquity and cosmic-ray acceleration}

\author[S. Banfi et al.]{S. Banfi$^{1,2}$\thanks{Contact e-mail: \href{serena.banfi2@unibo.it}{serena.banfi2@unibo.it}}, F. Vazza$^{1,2,3}$, D. Wittor$^{1,2,3}$
\\
$^{1}$Dipartimento di Fisica e Astronomia, Universit\`a di Bologna, Via Gobetti 92/3, 40121, Bologna, Italy\\
$^{2}$INAF, Via Gobetti 101, 40121 Bologna, Italy\\
$^{3}$Hamburger Sternwarte, Gojenbergsweg 112, 21029 Hamburg, Germany}

\date{}

\pubyear{2020}

\begin{document}
\label{firstpage}
\pagerange{\pageref{firstpage}--\pageref{lastpage}}
\maketitle

\begin{abstract}
Structure formation shocks are believed to be the largest accelerators of cosmic rays in the Universe. However, little is still known about their efficiency in accelerating relativistic electrons and protons as a function of their magnetization properties, i.e. of their magnetic field strength and topology. In this work, we analyzed both uniform and adaptive mesh resolution simulations of large-scale structures with the magnetohydrodynamical grid code \textsc{Enzo}, studying the dependence of shock obliquity with different realistic scenarios of cosmic magnetism.  We found that shock obliquities are more often perpendicular than what would be expected from a random three-dimensional distribution of vectors, and that this effect is particularly prominent in the proximity of filaments, due to the action of local shear motions. By coupling these results to recent works from particle-in-cell simulations, we estimated the flux of cosmic-ray protons in galaxy clusters, and showed that in principle the riddle of the missed detection of hadronic $\gamma$-ray emission by the \textit{Fermi}-LAT can be explained if only quasi-parallel shocks accelerate protons. On the other hand, for most of the cosmic web the acceleration of cosmic-ray electrons is still allowed, due to the abundance of quasi-perpendicular shocks. We discuss quantitative differences  between the analyzed models of magnetization of cosmic structures, which become more significant at low cosmic overdensities. 

\end{abstract}


\begingroup
\let\clearpage\relax
\endgroup
\newpage

\section{Introduction}
Shocks in the large-scale structure of the universe are the natural outcome of the accretion of cold, warm or hot gas onto galaxy clusters or of direct mergers between clusters \citep[e.g.][]{ry03,by08}. These processes convert a fraction of kinetic energy into thermal energy and into the amplification of magnetic fields and acceleration of cosmic rays (CRs) \citep[e.g.][for a recent review]{Bykov2019}. 
Through cosmological numerical simulations, we can estimate the energetics of CR associated to shocks: \citet[][]{mi00} provided the first attempt to simulate shock waves in the large-scale structure with an Eulerian approach and to derive their Mach number from jump conditions. Following works \citep[e.g.][]{ry03,pf06,va09shocks,2013MNRAS.428.1643P} found that, for the majority of shocks in the Universe, the kinetic energy is dissipated in internal shocks with low Mach numbers ($2\lesssim M\lesssim 4$), while shocks with Mach numbers up to $\sim1000$ are found in lower density environments (like the external accretion regions of structures) but they overall process little energy in the cosmic volume.  On the other hand, the acceleration of CRs by first order Fermi acceleration is expected to be mainly driven by strong shocks ($M\gtrsim5$) \citep[e.g.][]{ry03,kj07,va11comparison}. 

Radio observations of Mpc-sized synchrotron emission in galaxy clusters confirm the presence of diffuse magnetic fields and relativistic electrons associated with cluster merger shocks \citep[e.g. ``radio relics'', see][for reviews]{fe08,fe12,2019SSRv..215...16V}, while at the same time the lack of hadronic $\gamma$-ray detection by the \textit{Large Area Telescope} (LAT) on board of the \textit{Fermi} satellite \citep[e.g.][]{ack10,arl12,fermi14} has set stringent upper limits on the content of CR protons in galaxy clusters ($\lesssim 1\ \%$ of the thermal gas energy), which also can be used to set very low upper limits on the allowed CR acceleration efficiency of structure formation shocks  \citep{va14relics,bj14}.

Several decades of theoretical works suggest that each kind of particles undergoes different levels of acceleration as a function of plasma parameters and of the pre-existing magnetic field topology, but the mechanism that drives this process is still under debate \citep[e.g.][and references therein]{Bykov2019}. CRs typically gain energy by crossing the shock front multiple times through a first order Fermi mechanism called diffusive shock acceleration \citep[DSA, ][]{be78}, which produces a power-law distribution of energetic particles. However, unlike protons, electrons need to be pre-accelerated in order for their small gyro-radius to become comparable to the width of the shock front and effectively enter the DSA regime:  particle-in-cell (PIC) simulations by \citep[e.g.][]{2014ApJ...783...91C,guo14} suggest that shock drift acceleration \citep[SDA, e.g.][and references therein]{2011ApJ...742...47M} could be an efficient way for these particles to be pre-accelerated by drifting along magnetic field lines down the shock front.

\begin{figure*}
\includegraphics[scale=1,clip,trim={0.3cm 0.5cm 0.3cm 0.5cm}]{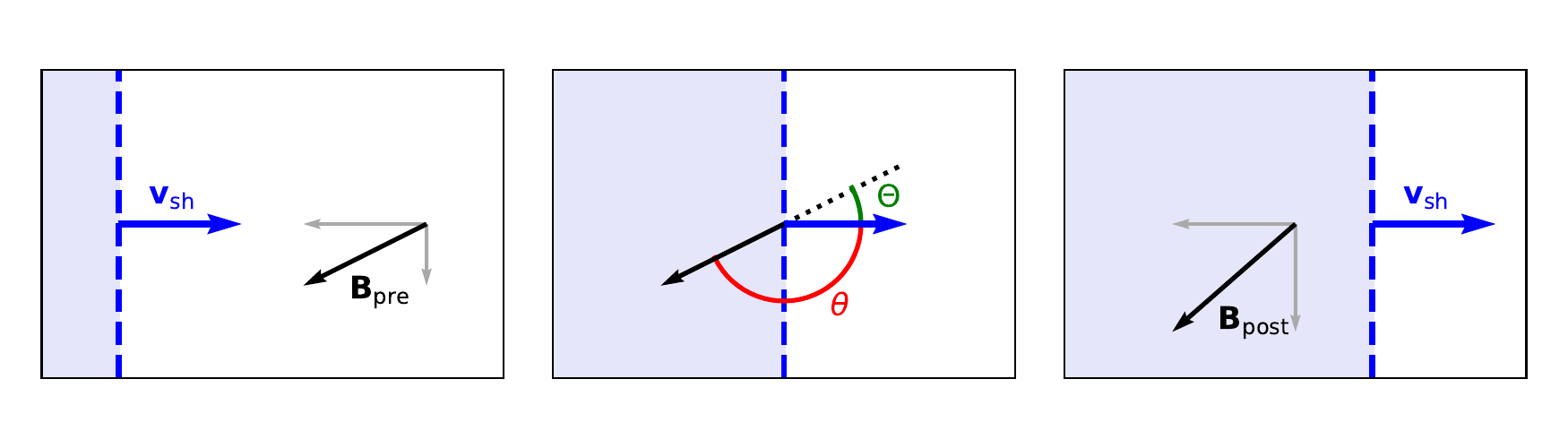}
\caption{The Figure shows the two-dimensional projection of the process of shock crossing on magnetized plasma in the rest frame of the pre-shock.  The left panel shows the pre-shock configuration; the central panel represents the obliquity $\theta$, i.e. the angle formed by the shock propagation direction and the pre-shock magnetic field, and $\Theta=90^{\circ}-\left|90^{\circ}-\theta\right|$ (for acute angles, $\theta\equiv\Theta$); the right panel shows the magnetic field modification after shock crossing, i.e. the amplification of its perpendicular component.}
\label{fig:thetaTheta}
\end{figure*}

A usually underlooked aspect of particle acceleration from cosmic shocks is the role of shock obliquity $\theta$, defined as the angle between the shock normal and the up-stream magnetic field vector (see Figure \ref{fig:thetaTheta}). PIC simulations have indeed established the dependence of the acceleration on the shock's Mach number and obliquity: as a consequence CR electrons have been shown to be more easily accelerated by quasi-perpendicular ($45^{\circ}<\theta<135^{\circ}$, \citealt{guo14}) rather than quasi-parallel ($\theta <45^{\circ}$ or $\theta >135^{\circ}$) shocks, while the opposite has been found for protons \citep{2014ApJ...783...91C}. However, magnetohydrodynamical (MHD) simulations are necessary to study the conditions that lead to a certain orientation of the magnetic field where shocks occur and to the possible prevalence of some obliquities over others.

The link between obliquity and acceleration efficiency may also be a viable explanation for the missing $\gamma$-ray detection from galaxy clusters, considering that the distribution of random angles in a three-dimensional space is $\propto\sin\theta$, which means that it is peaked at perpendicular shocks. Simulations by \citet[][]{wi17} have shown that the obliquity distribution of shocks in galaxy clusters progressively becomes even more concentrated towards $90^{\circ}$\ as a result of the passage of several merger shock waves in the lifetime of clusters. 
They also estimated that, if the acceleration of CR protons is limited to shocks with $\theta <50^{\circ}$, the hadronic $\gamma$-ray emission produced by DSA gets much reduced and the tension with \textit{Fermi} limits is alleviated, even if not entirely solved. 
More recent work from \citet[][]{Ha2019} has shown that in simulated galaxy clusters the amount of kinetic energy flux dissipated by quasi-parallel shocks and  transferred to CR protons is $\sim 10^{-4}$, assuming a DSA model with more recent efficiencies derived in \citet[][]{Ryu2019}. In this case, the obtained $\gamma$-ray emissions are in line with \textit{Fermi}'s constraints. 
This picture has been recently confirmed by cosmological MHD simulations by \citet{wittor20}.\\

However, we notice that \citet[][]{Ha2019} did not use MHD simulations of large-scale structures (but rather a simpler approach involving the evolution of passive magnetic fields via the induction equation), while \citet{wi17} did not focus on the properties of shock acceleration at the scale of filaments, which are expected to be a major contributor to the total mass content of galaxy clusters. Moreover, the above works were limited to explore the dependence of shock acceleration on the magnetic fields generated by single possible scenarios for the origin of extragalactic magnetic fields, whose origin is still unclear \citep[e.g.][]{donn09,va17cqg}.

Our work expands on the above points, using new MHD simulations tailored to determine the typical obliquity of cosmic shocks in the acceleration of relativistic particles across the cosmic environment, its potential dependence on magnetogenesis and its effect on the acceleration efficiency of electrons and protons.

The paper is structured as follows. In Section \ref{sec:meth}, we describe the computational setup for the simulations and we outline the approach used to find and characterize shocks. In Section \ref{sec:res}, we present our results for obliquity and CR acceleration estimates. In Section \ref{sec:obs}, we describe the implications of our results on observations. In Section \ref{sec:disc}, we discuss the validity and limitations of our analysis. Finally, Section \ref{sec:con} contains a brief summary and conclusions. In the Appendix, we clarify some details about the analysis.

\section{Methods}
\label{sec:meth}
\subsection{Simulations}
\label{sec:sim}

We simulated the formation of cosmic structures with the Eulerian cosmological magnetohydrodynamical code \textsc{Enzo} \citep[][]{enzo14}, which couples an N-body particle-mesh solver for dark matter \citep{he88} with an adaptive mesh refinement (AMR) method  for the baryonic matter \citep{bc89}. We adopted a piecewise linear method (PLM) \citep{1985JCoPh..59..264C}, a reconstruction technique in which fluxes are computed using the Harten-Lax-Van Leer (HLL) approximate Riemann solver, and used time integration based on the total variation diminishing (TVD) second-order Runge-Kutta (RK) scheme \citep{1988JCoPh..77..439S}.
We used the Dedner cleaning MHD solver \citep{ded02} to keep the divergence of the simulated magnetic field as small as possible. This method has been tested multiple times in the literature, 
showing that despite the relatively large rate of dissipation introduced by its ``cleaning waves'', it always converges to the correct solution as resolution is increased,  at variance with other possible ``divergence cleaning'' methods \citep[e.g.][]{2013MNRAS.428...13S,2016MNRAS.455...51H,2016MNRAS.461.1260T}.  We refer the reader to more recent reviews for a broader discussion of the resolution and accuracy of different MHD schemes  in properly resolving the dynamo in cosmological simulations \citep{review_dynamo}.

In this work, we present the analysis of two kinds of simulations (described in more detail in the next two Sections): a suite of runs employing a fixed spatial/mass resolution ($\approx 83\ \mathrm{kpc/cell}$ comoving) to simulate the cosmic web on a representative cosmic volume and for different scenarios of the origin of magnetic fields, and single cluster re-simulations using nested initial conditions, which allow us to study magnetic field topology at high resolution ($\approx 25\ \mathrm{kpc/cell}$ comoving) for a specific scenario of cosmic magnetism. 

\subsection{Static grid simulations}
\label{sec:chronos}
We simulated a volume of $\approx(85\ \mathrm{Mpc})^3$ (comoving) sampled with a static grid of $1024^3$ cells: the decision to neglect AMR allows us to maintain a resolved description of magnetic fields even in low-density regions. These datasets are extracted from the ``Chronos++'' suite{\footnote{\url{http://cosmosimfrazza.myfreesites.net/the\_magnetic\_cosmic\_web}}}, which includes a total of $24$ simulations aimed at exploring different possible scenarios concerning the origin and evolution of magnetic fields in the cosmic web environment \citep{va17cqg}. Here, we focus on four of the most realistic models, characterized by  relevant variations of the topology of magnetic fields, which is very interesting for our study:

\begin{enumerate}
 \item \textit{``baseline''}: non-radiative run with a primordial uniform volume-filling comoving magnetic field $B_0=1\ \mathrm{nG}$ at the beginning of the simulation;
  \item \textit{``Z''}: non-radiative run with a primordial magnetic field oriented perpendicularly to the velocity vector, as in \citet{va17cqg}; in order to ensure that $\nabla \cdot \mathbfit{B} \equiv 0$ at the beginning, the starting magnetic field vectors were initialized  perpendicularly to the three-dimensional gas velocity field computed with the Zeldovich approximation \citep[e.g.][] {do08}, so that a purely solenoidal initial field is produced and thus enforce the $\nabla \cdot \mathbfit{B} \equiv 0$ condition construction. The r.m.s. values of each component generated with the Zeldovich approximation are renormalized within the cosmic volume, in order to match the same level of seed magnetic field as in the baseline primordial run, i.e.  $\sqrt{\langle B^2\rangle} = B_0$; 
 \item \textit{``DYN5''}: non-radiative run with sub-grid dynamo magnetic field amplification. The small-scale dynamo amplification of a weak seed field of primordial origin  ($B_0 = 10^{-9}\ \mathrm{nG}$ comoving) is computed at run-time. In this model, the dissipation of solenoidal turbulence into magnetic field amplification is estimated at run-time  based on the measured gas velocity vorticity in the cells, with a flow Mach number-dependent efficiency derived from \citet{fed14}. This model attempts to bracket the possible residual amount of dynamo amplification which can be achieved in low density environments as suggested by other works \citep[e.g.][]{ry08}, but is  lost due to finite resolution effects \citep[see][for more details]{va17cqg};
 \item \textit{``CSFBH2''}: radiative run with an initial background magnetic field $B_0=10^{-10}\ \mathrm{nG}$ (comoving) including gas cooling, chemistry, star formation, thermal/magnetic feedback from stellar activity and active galactic nuclei (AGN). In this model, supermassive black hole (SMBH) particles are initially added to the simulation at $z=4$ at the center of massive halos as ``seeds'', with an initial mass of $M_{\mathrm{BH,0}}=10^4\ \mathrm{M_{\odot}}$ \citep[][]{2011ApJ...738...54K}. From that moment on, they accrete matter from the gas distribution in the grid based on the  spherical Bondi-Hoyle formula, assuming a fixed $\sim 0.01\ \mathrm{M_{\odot}/yr}$ accretion rate and a ``boost'' factor to the mass growth rate of SMBH ($\alpha_{\mathrm{Bondi}} =1000$, calibrated to balance the unresolvable gas clumping in the innermost accretion regions). Star forming particles are also generated on the fly based on the \citet{2003ApJ...590L...1K} star formation recipe, which also accounts for the thermal feedback of star formation winds onto the surrounding gas.  In this simulation we coupled the \textsc{Enzo} thermal feedback from SMBH and from star forming particles to the injection of additional magnetic energy via bipolar jets, with efficiencies $\epsilon_{\mathrm{SF,b}}=10 \ \%$ and $\epsilon_{\mathrm{BH,b}}=1 \ \%$ for the star formation and SMBH respectively, referred to the corresponding  $\dot{M} c^2$ energy accreted in the two processes. While all above prescriptions obviously represent a gross oversimplification of the very complex physics behind star formation and black hole evolution, such ad-hoc models  have been calibrated and chosen out of the larger sample of simulations tested in \citet{va17cqg} as they provide a good match to the observed cosmic average star formation rate as well as to observed galaxy cluster scaling relations. Our main purpose in using this model is to have a realistic representation of the impact of galaxy formation physics on the $\gtrsim 100\ \mathrm{kpc}$ scales which are relevant for our study of cosmic magnetism and structure formation shocks. 
 
\end{enumerate}

The cosmological parameters were chosen accordingly to a $\Lambda$CDM cosmology: $H_0=67.8\ \mathrm{km\ s^{-1}\ Mpc^{-1}}$, $\Omega_{\mathrm{b}}=0.0468$, $\Omega_{\mathrm{m}}=0.308$, $\Omega_{\mathrm{\Lambda}}=0.692$ and $\sigma_8=0.815$ \citep[][]{2016A&A...594A..13P}.
In the following, we will refer to these simulations as to the ``Chronos'' runs. 

\subsection{Nested grid simulations}
\label{sec:sp}
 To study the evolution of obliquity and magnetic fields around galaxy clusters at a resolution more comparable to the one of radio observations,  we also examined six simulations from the ``San Pedro'' suite\footnote{\url{https://dnswttr.github.io/proj_sanpedro.html}} \citep[][]{wittor20}. This set of simulations uses nested grids to assure a uniform resolution even at the most refined level. Each cluster was extracted from the box with a root-grid size of $(207 \ \mathrm{Mpc})^3$ sampled with $256^3$ cells. A region of $\approx (6.5 \ \mathrm{Mpc})^3-(9.8 \ \mathrm{Mpc})^3$ was further refined using five levels, i.e. $2^5$ refinements, of nested grids. The procedure guarantees a uniform resolution of $\approx 25 \ \mathrm{kpc}$ comoving around the clusters, from the beginning to the end of the run. The sizes of the nested regions ensure that their volume is at least $3.5^3$ times larger than the volume enclosed in $r_{200}^3$\footnote{$r_{200}$ is defined as the radial distance from the cluster center inside which the mean density is $200$ times the critical density.}.  The initialization of the nested grids was performed using \textsc{MUSIC} \citep{music}. In each simulation, the initial magnetic field is uniform and takes a value of $0.1 \ \mathrm{nG}$ (comoving) in each direction. The six clusters used in this work are characterised by different evolutionary stages, ranging from pre-merger, over actively merging, to post-merger.
 
We note that these simulations use slightly different cosmological parameters than the Chronos runs (see Section \ref{sec:chronos}). The parameters of these runs are based on the latest results from the Planck-collaboration \citep[i.e.][]{PlanckVI2018}: $H_0=67.66\ \mathrm{km\ s^{-1}\ Mpc^{-1}}$, $\Omega_{\mathrm{b}}=0.04831914$, $\Omega_{\mathrm{m}}=0.3111$, $\Omega_{\mathrm{\Lambda}}=0.6889$ and $\sigma_8=0.8102$. Furthermore, for code stability issues at the fixed refinement regions in this case we used the somewhat more diffuse Local Lax-Friedrichs (LLF) Riemann solver to compute the fluxes in the PLM. In the following, we will refer to the set of nested simulations as to the ``San Pedro'' runs.

\subsection{Shock finder}
\label{sec:shfin}
The shock finding method is applied in post-processing and it is based on \citet{ry03}, with the changes explained in \citet{va09shocks}. It allows to determine the Mach number $M$ of a shock from the velocity jump $\Delta v \leq 0$ between pre-shock and post-shock cells:
\begin{equation}
\label{vj}
    \Delta v=\frac{3}{4} v_s \frac{1-M^2}{M^2},
\end{equation}
where $v_s=Mc_s$ and $c_s$ is the sound speed of the pre-shock cell. The procedure is the following:
\begin{enumerate}
    \item candidate shocked cells are selected with the constraint on the three-dimensional velocity divergence $\nabla\cdot \mathbfit{v}<0$;
    \item for each Cartesian direction, the pre-shock (post-shock) cell is identified as the one with the minimum (maximum) temperature at a distance $\Delta x$ from the candidate cell. We investigated different values for $\Delta x=1,2,3$, which serves as a stencil for the computation of the Mach number via jump conditions. This is motivated by the fact that numerical shocks (especially if they are oblique with respect to the grid) are not an ideal discontinuity but are typically broadened across a few cells: hence, the jump conditions must be computed over a large enough distance.    %
    Our tests have shown that the optimal choice here is $\Delta x=1$, since larger steps would include the contribution of contaminating flows unrelated to the shock, consistently with previous work \citep[][]{va09shocks};
    
    \item the shock Mach number is given by Equation \ref{vj} for each of the three dimensions, preceded by a sign indicating the orientation of the velocity jump: in case multiple contiguous cells are flagged as shocked along the same direction, those with the lowest absolute Mach number are discarded;
    \item the final Mach number is calculated as $M=\sqrt{M_x^2+M_y^2+M_z^2}$ and it is assigned to the post-shock cell;
    \item only shocks above a certain threshold of Mach number (which we calibrated depending on the resolution of the specific simulation, see Section \ref{sec:res_sp}) are considered in order to prevent spurious identification in the complex gas flows of galaxy clusters \citep[e.g.][]{va09shocks}.
    
\end{enumerate}

The three components of the Mach number give the direction along which the shock propagates, i.e. the normal to the shock front. %

\subsection{Shock obliquity}
\label{sec:obl}

The obliquity is computed as the angle between the propagation direction and the magnetic field $\mathbfit{B}$ in the pre-shock cell:
\begin{equation}
    \theta=\arccos{\left(\frac{M_x\cdot B_x+M_y\cdot B_y+M_z\cdot B_z}{M\ B}\right)}.
\end{equation}
and it ranges from $0^{\circ}$ to $180^{\circ}$. In the following analysis we will mostly refer to the quantity defined as %
$\Theta=90^{\circ}-\left|90^{\circ}-\theta\right|$, which ranges from $0^{\circ}$ to $90^{\circ}$: this way, for example, shocks with $\theta=20^{\circ}$ and $\theta=160^{\circ}$ are considered equivalent in terms of CR acceleration efficiency (see Figure \ref{fig:thetaTheta}).

\begin{figure}
\centering
\includegraphics[width=0.5\columnwidth,clip,trim={6cm 4cm 5cm 6cm}]{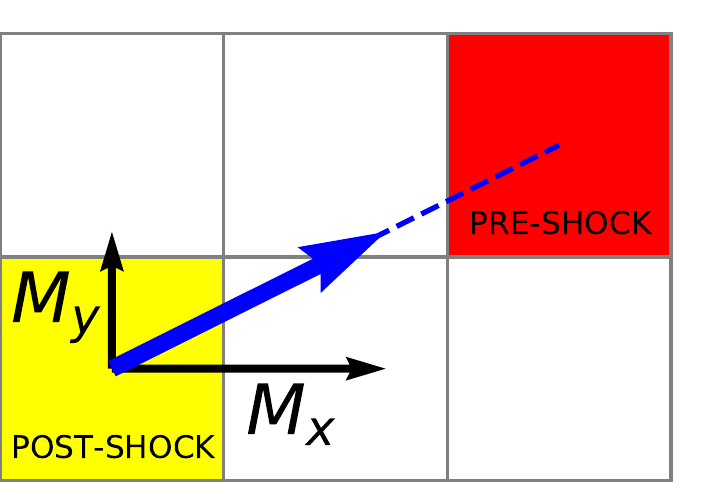}
\caption{Sketch of the adopted procedure to identify the pre-shock region (red) of a post-shock cell (yellow) in our simulation, under the prior that the numerical shock structure is always spread over at least three cells  (here a simple two-dimensional case is given for simplicity). }
\label{fig:presh}
\end{figure}

During step (iii) of the shock finding procedure, for each shock we identify three pre-shock cells, one for each direction, in order to  compute the Mach number components $M_x$, $M_y$ and $M_z$ from jump conditions. Knowing the up-stream magnetic field orientation is necessary to measure the shock obliquity, so we need to identify a single cell as the pre-shock. Determining the proper pre-shock cell for each identified shocked cell is not always a trivial operation, as in converging flows or shocks with complex pre-shock conditions this operation is affected by some level of uncertainty, and numerical shocks are typically spread over (at least) three cells. Hence, to locate the pre-shock of a given shocked cell, we have to move two cells away, along the direction suggested by the measured Mach number. 
Figure \ref{fig:presh} gives the example of the reconstruction of post-shock and pre-shock cells for an oblique shock (limited to the two-dimensional case for simplicity). 
In this analysis, we choose the pre-shock cell as the one that is bound to be crossed by the shock at the following timestep in the simulation. In practice, a cell is tagged as a pre-shock cell, if a) it is located in the up-stream of the shock, b) it lies along the shock normal and c) it is located at a distance of two grid cells from the post-shock cell. 

This approach has provided reasonable identification of pre-shock cells, which can be we visually checked in the case of large-scale shocks, such as those surrounding cosmic filaments or galaxy clusters. Another way to determine the reliability of this method is to evaluate the conservation of the component of the magnetic field parallel to the putative shock normal, as required by idealized MHD: we discuss this issue in more detail in Appendix \ref{app:parall}.

Figure \ref{fig:zoom} gives an example of the distribution of shocks around a galaxy cluster (and around a zoomed filament) as well as of their obliquity, as measured by our method. Shocks are distributed around the cluster and the filaments with a large range of angles, yet with a clear predominance of quasi-perpendicular geometries ($\Theta > 45^{\circ}$). A higher resolution view of the distribution of shock obliquities around clusters and during their formation will be given in more detail in Section \ref{sec:res_sp}.

\begin{figure*}
\includegraphics[scale=1,clip,trim={0.8cm 0 0.6cm 0}]{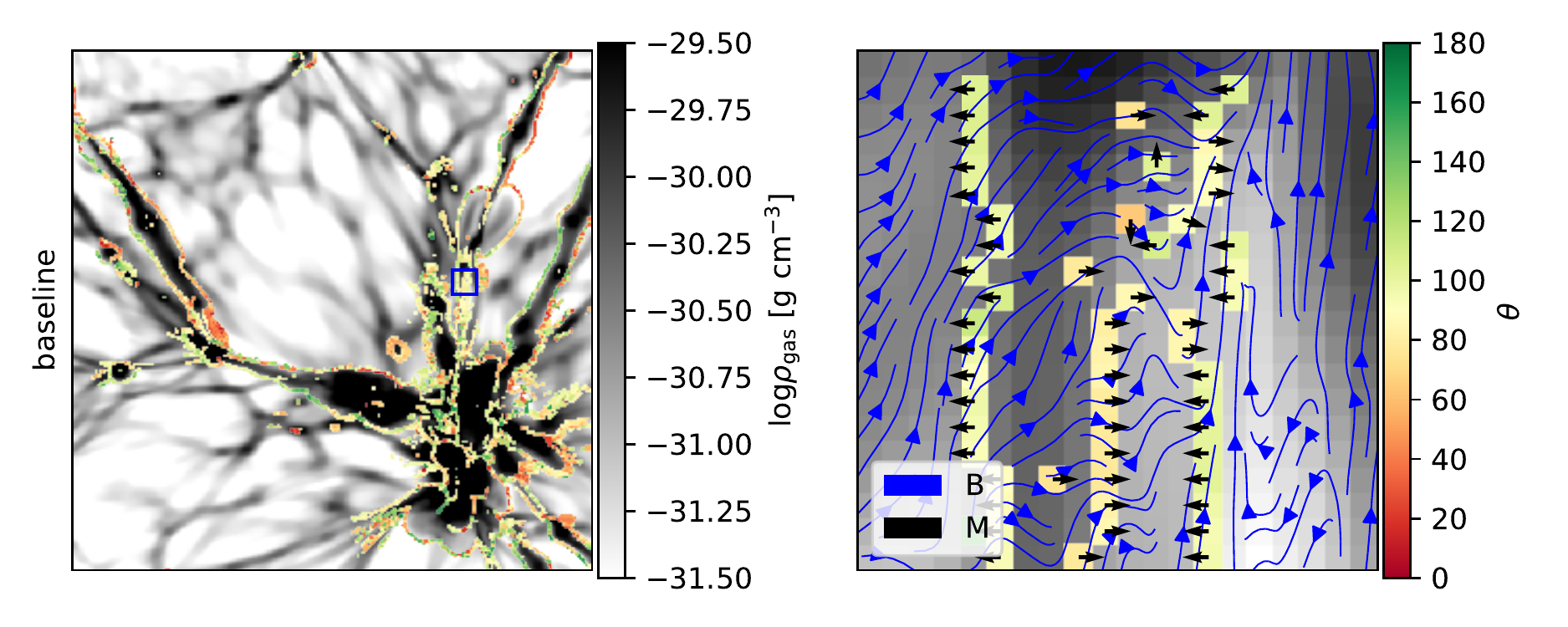}
\caption{Density slice of the baseline simulation at redshift $z=0$, with shocks color-coded by obliquity: the highlighted square on the left is zoomed on the right, where the arrows represent the projected magnetic field and Mach number. The box length is $\approx 41\ \mathrm{Mpc}$ on the left and $\approx 1.6\ \mathrm{Mpc}$ on the right.}
\label{fig:zoom}
\end{figure*}

\section{Results}
\label{sec:res}
\subsection{Analysis of full cosmic volumes}
\label{res_chronos}
\subsubsection{Thermal, magnetic and shock properties at $z=0$}
\label{sec:chpr}

\begin{figure}
\includegraphics[width=\columnwidth,clip,trim={0.5cm 0.5cm 0.5cm 0.5cm}]{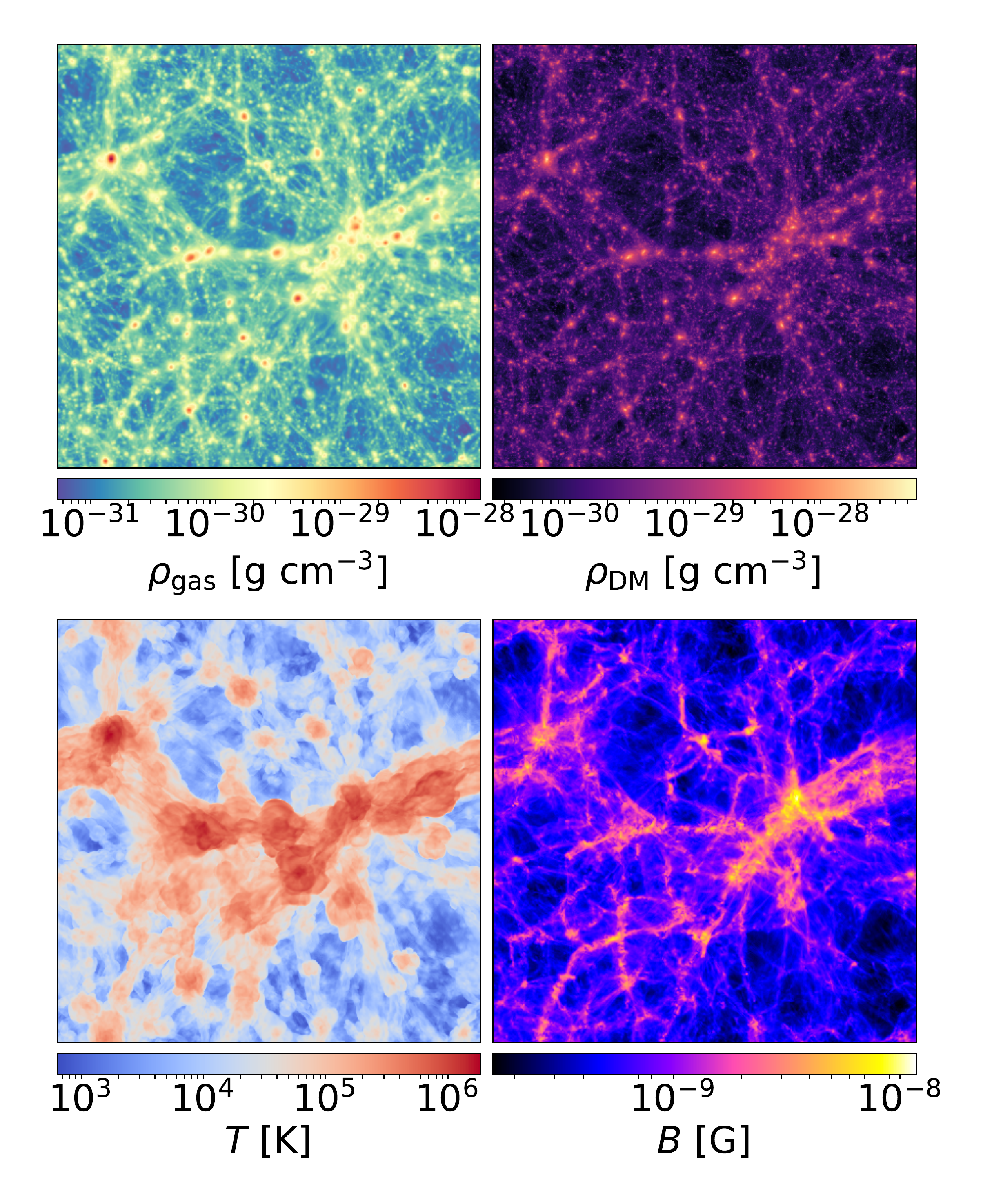}
\caption{Maps of average gas density, dark matter density, temperature and magnetic field volume-weighted along the line of sight for the baseline simulation at $z=0$. The box length is $\approx 85\ \mathrm{Mpc}$.}
\label{fig:maps}
\end{figure}

First, we analyze the final properties of shocks and magnetic fields in our  four different Chronos unigrid runs, described in Section \ref{sec:chronos}. In Figure \ref{fig:maps}, we show the projected maps of gas
density, dark matter density, gas temperature and magnetic field strength integrated along the line of sight, for our baseline simulation at $z=0$. The maps show the usual clustering of cosmic matter into galaxy groups and galaxy clusters ($\sim 10^6\ \mathrm{K}$ in the projected temperature map), cosmic filaments ($\sim 10^4-10^5\ \mathrm{K}$) and voids. Following from the density contrast formed within structures, as well as partially from the dynamo amplification within the largest halos, the projected magnetic field varies over two orders of magnitude from voids to halos. In reality, the range of difference in the three-dimensional grid is even higher, and the magnetic field within halos can reach $\sim 0.1-1\ \mathrm{\mu G}$ (e.g. Figure \ref{fig:phase} below). 
For other visualizations of the distribution of magnetic fields in these runs we refer the reader to \citet{va17cqg} and \citet{gv19}. 
From the combination of the above trends we can expect that the thermal gas energy and the magnetic energy (and possibly their ratio) can vary very significantly across the simulated volume.

\begin{figure}
\includegraphics[width=\columnwidth,clip,trim={0.5cm 0.5cm 0.5cm 0.5cm}]{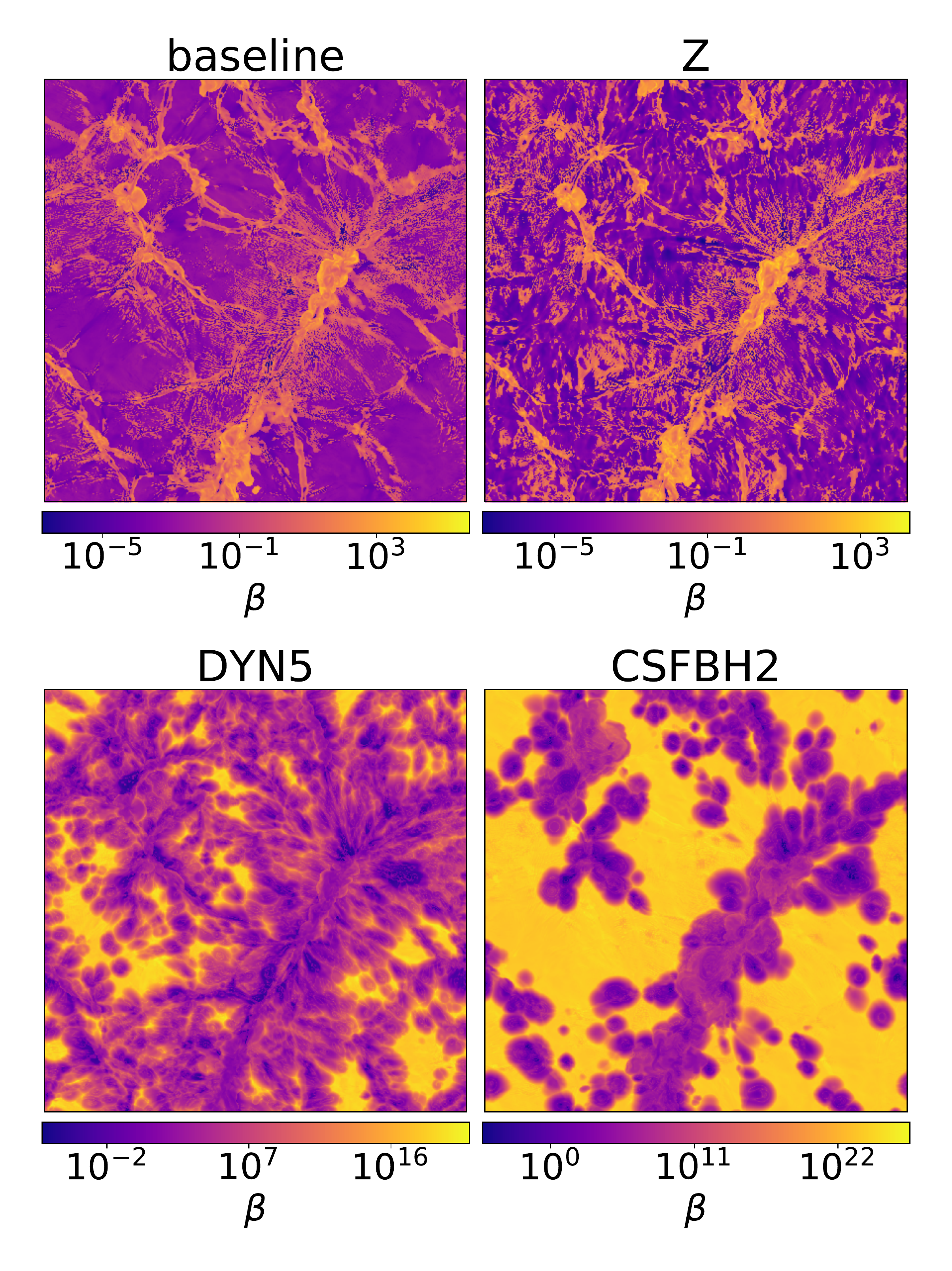}
\caption{Plasma $\beta$ of a slice of volume with thickness of $83\ \mathrm{kpc}$ for the four Chronos simulations at $z=0$. The box length is $\approx 85\ \mathrm{Mpc}$.}
\label{fig:beta}
\end{figure}

In particular, the relative importance of the thermal component with respect to the magnetic component in a magnetized plasma is parametrized by the quantity $\beta=n k_{\mathrm{B}} T/(B^2/8\pi)$, defined as the ratio of the thermal pressure over the magnetic pressure. In Figure \ref{fig:beta}, we show the values of $\beta$ in the same slice of volume for the four simulations. In all runs, there is a considerable difference between the values of $\beta$ in virialized structures and in voids, but the trends are opposite for the first two and last two runs, and are mostly driven by the different scenarios for magnetogenesis. The baseline and Z simulations are overall characterized by low values of $\beta$, due to their stronger seed magnetic field. In such cases, $\beta$ is highest in clusters and filaments, owing to the larger value of gas pressure there. On the other hand, in runs with weak seed magnetic fields (DYN5 and CSFBH2) the thermal pressure in voids always dominates over the very weak magnetic pressure ($\beta\gg1$), while dynamo and stellar evolution amplify magnetic fields to $\beta \sim 1-10^2$, only within dense structures.  
Therefore, in our scenarios there are several environments in which the effect of magnetic pressure is not negligible compared to the thermal pressure. However, it shall also be noticed that even where $\beta \sim 1$, in the cosmic volume the kinetic ram pressure is always dominant, due to the typically large ($\sim 10^2-10^3\  \mathrm{km\ s^{-1}}$) infall motions induced by accretions: hence, in most cases the magnetic fields are still being passively advected in the cosmic volume.

\begin{figure}
\includegraphics[scale=1,clip,trim={0.2cm 0 0.3cm 0}]{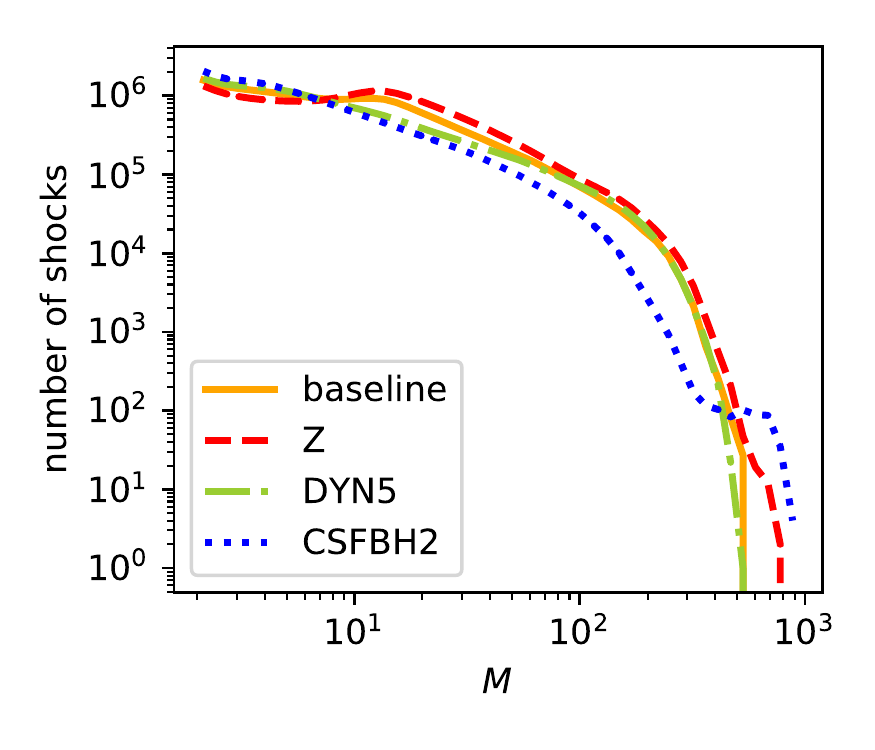}
\caption{Number of occurring shocks per intervals of Mach number for the four Chronos simulations  at $z=0$.}
\label{fig:mach}
\end{figure}

We analyzed shocked cells in the simulated volume and we selected only shocks with $M>2$, since weaker shocks are expected to be unable to accelerate CRs \citep{Ha2018protons}: identified shocks in all runs correspond to $\approx 2\ \%$ of the total number of cells. %
In Figure \ref{fig:mach}, we show the distribution of shock Mach numbers in the entire volume. The shape is consistent with previous results from the literature \citep[e.g.][]{ry03}, i.e. cosmological shocks belong to two distinct populations, external and internal shocks, identifying shocks affecting gas with pre-shock temperature $\lesssim10^4\ \mathrm{K}$ or $\gtrsim 10^4\ \mathrm{K}$ respectively, the latter meaning that the material had already been previously shocked. This division can be observed in Figure \ref{fig:mach}, where the bump at $M\approx 20$ marks the separation between weaker internal shocks, associated to mergers, from stronger external shocks surrounding filaments. The distribution of CSFBH2 deviates from the others: at low $M$, a significantly larger number of shocks is generated in high-temperature regions due to AGN feedback, consistently to previous works \citep[e.g.][]{ka07,va13feedback}.  On the other hand at high $M$, due to heating effect of reionization modelled in CSFBH2, the gas temperature is increased to $\sim 10^4\ \mathrm{K}$, preventing the sound speed to be as low as in the other runs.

\begin{figure*}
\includegraphics[scale=1,clip,trim={0 0.8cm 6cm 0cm}]{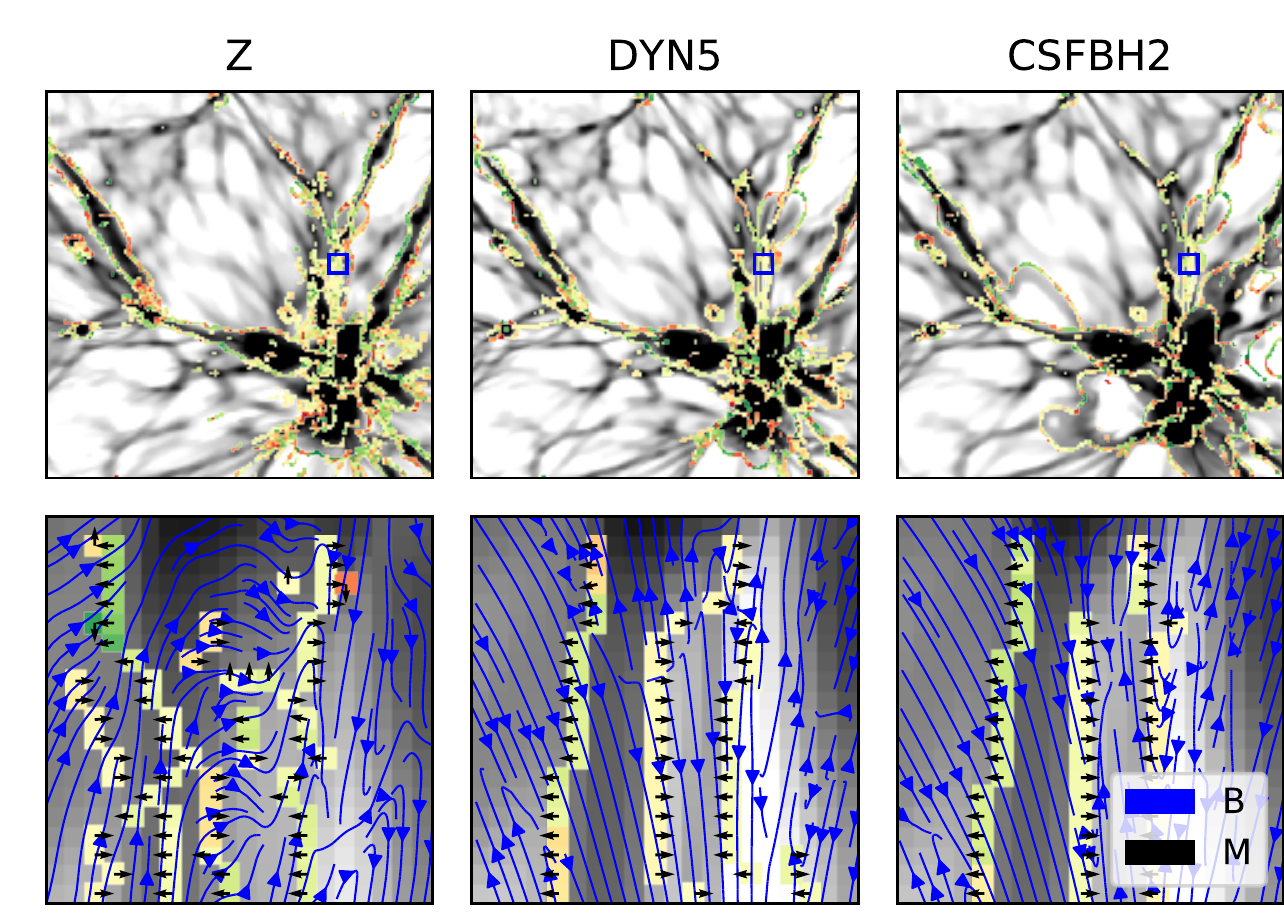}
\includegraphics[scale=1,clip,trim={16.5cm 0.8cm 0 0}]{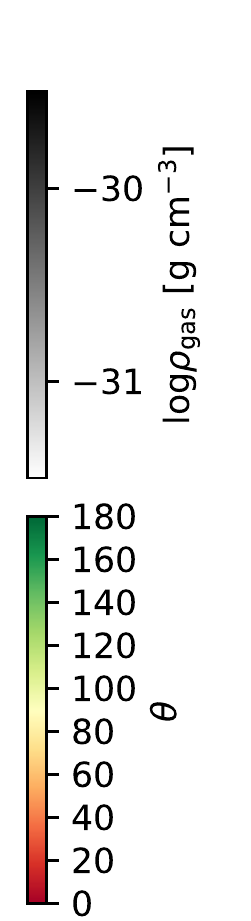}
\caption{Same as Figure \ref{fig:zoom} for the remaining Chronos simulations.}
\label{fig:zoom234}
\end{figure*}

\begin{figure}
\includegraphics[scale=1,clip,trim={0.3cm 0.2cm 0.3cm 0.2cm}]{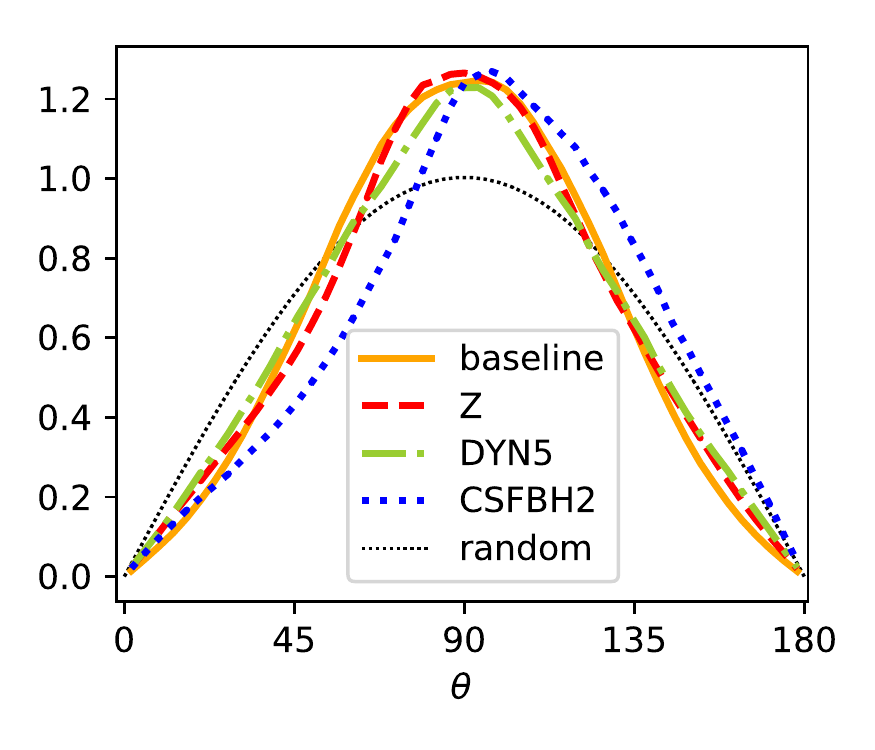}
\caption{Number distribution of shocks per intervals of obliquity for the four Chronos simulations at $z=0$, normalized to the random distribution.}
\label{fig:oblhist}
\end{figure}

We computed the obliquity for each shock (as in Section \ref{sec:obl}): Figure \ref{fig:zoom234}, analogously to Figure \ref{fig:zoom}, shows the distribution of shocks and the corresponding obliquity for a slice of the simulated volume for the remaining three runs.
The density slice remains mostly unvaried from one run to another, as well as the location of the zoomed filament: however, the magnetic field topology differs, thus affecting the obliquity.
We then investigated the deviation of obliquity from the distribution of angles expected from random vectors in space, which is $\propto \sin\theta$. Figure \ref{fig:oblhist} shows the number of identified shocks per intervals of $\theta$. The curves have a higher peak at perpendicular shocks with respect to the random distribution: previously-shocked gas hosts shocks that are on average more perpendicular than in the random distribution, likely due to the compression of the perpendicular component of the magnetic field after shock crossing, consistently to what \citet{wi17} found. The CSFBH2 run shows a skewed distribution that may be attributed to the few bursts of star and/or AGN feedback which still occur at low redshift, and that can thus vary from snapshot to snapshot.

\begin{figure}
\includegraphics[clip,trim={0.3cm 0.3cm 0.3cm 0}]{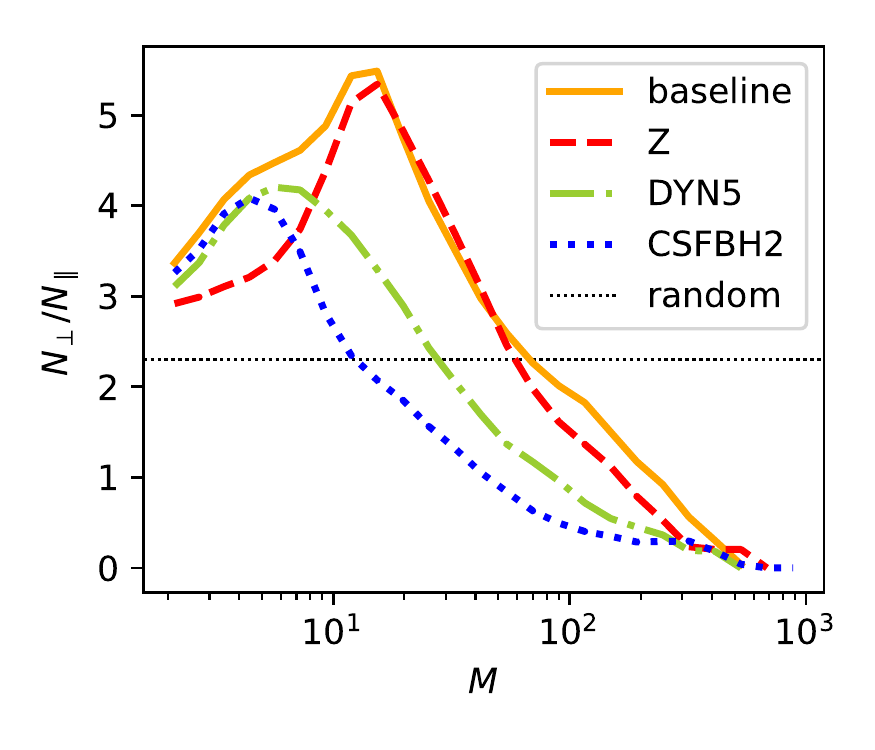}
\caption{Ratio of number of quasi-perpendicular shocks and number of quasi-parallel shocks per Mach number intervals for the four Chronos simulations at $z=0$, compared to the ratio expected for randomly distributed obliquities.}
\label{fig:npp}
\end{figure}

Figure \ref{fig:npp} shows the excess of quasi-perpendicular shocks ($\Theta>45^{\circ}$, capable of accelerating electrons) with respect to quasi-parallel shocks ($\Theta<45^{\circ}$, capable of accelerating protons) as a function of Mach number. As it can be seen in Figure \ref{fig:oblhist}, for vectors randomly distributed in space a perpendicular configuration is more probable than a parallel one. In particular, a random configuration would return a value of $N_{\perp}/N_{\parallel}\approx 2.3$ integrated over the entire distribution of angles. Therefore,  for weak shocks, quasi-perpendicular geometries are way more frequent than by chance, and thus the (magneto)hydrodynamics of gas is playing a role in aligning magnetic field vectors with the shock surface. We will discuss on the physical interpretation of this phenomenon in more detail in Section \ref{sec:res_sp} (see also Appendix \ref{app:xi}). The opposite trend is found for $M\gtrsim 50$ shocks: at least in part, this can be ascribed to numerical problems. Shocks in this regime are not energetically relevant and are typically confined in very low density environment; the numerical uncertainties related to the modelling of shock obliquity in this regime are discussed in detail in Appendix \ref{app:waves}. 

\begin{figure}
\includegraphics[clip,trim={0.3cm 0.3cm 0.3cm 0}]{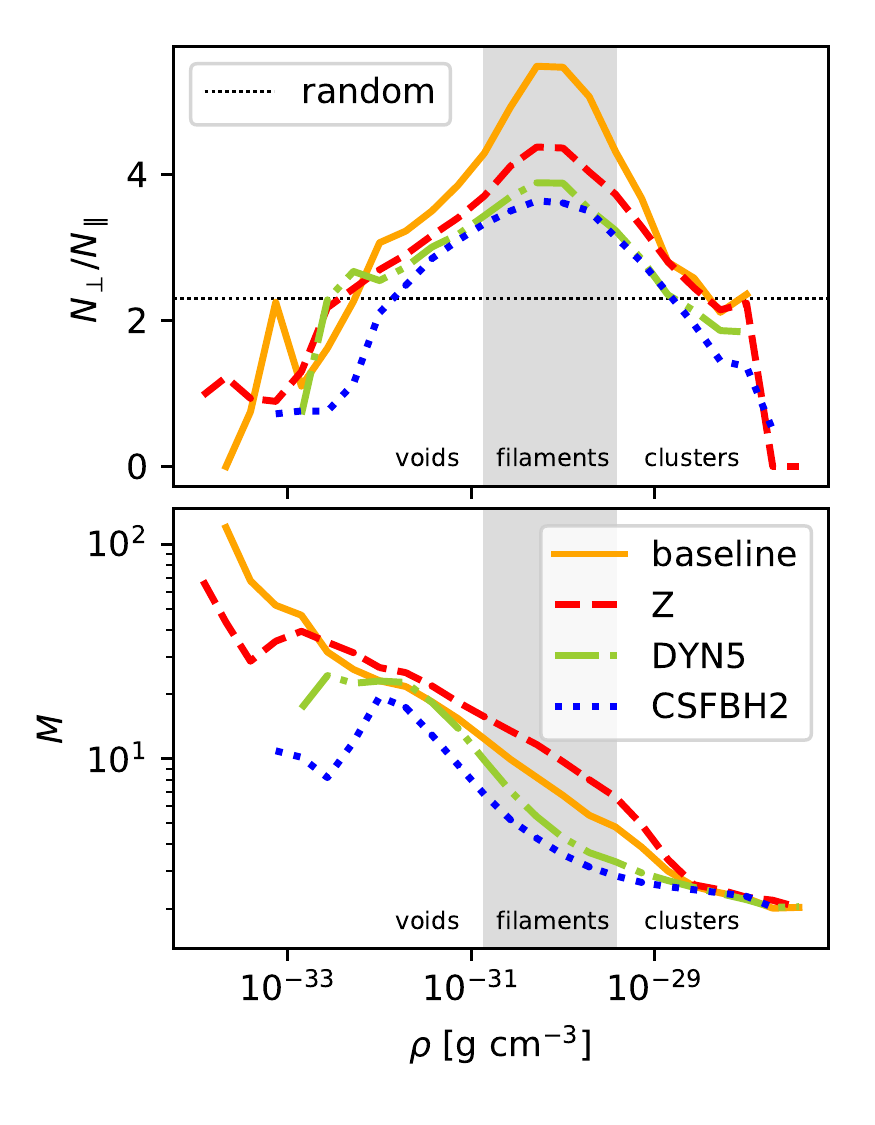}
\caption{Top panel: ratio of number of quasi-perpendicular shocks and number of quasi-parallel shocks per density intervals for the four Chronos simulations at $z=0$, compared to the ratio expected for randomly distributed obliquities. Bottom panel: mean Mach number per density intervals for the four Chronos simulations. In both panels the range in which the overabundance of perpendicular shocks is largest is highlighted in grey: this corresponds roughly to the regions hosting filaments.}
\label{fig:npp_d}
\end{figure}

The highest overabundance of perpendicular shocks is found at $M \approx 10$, with the exact location of the peak changing from one run to another. By computing the same ratio as a function of pre-shock density, we constrain the peak to be located at $\rho \approx 10^{-30}\ \mathrm{g\ cm^{-3}}$, independently of the specific re-simulation (Figure \ref{fig:npp_d}). This means that these shocks, which are more perpendicular than average, are likely to be generated in the same cosmic environment, regardless of the specific model for magnetism. Based on the density range, we can constrain these shocks to be associated with filaments of the cosmic web, which is also supported by visual inspection. However, the same density interval does not correspond to the same Mach number interval in the four runs: in CSFBH2 this is explained by the rise in temperature due to reionization, which limits the strength of shocks. Also in DYN5 we measure a larger temperature than in the other non-radiative runs, leading to slightly weaker shocks \citep[][]{gv19}.
Filaments of the cosmic web are thus an environment where the acceleration of CRs mostly happens via quasi-perpendicular shocks, with implications on the injection and evolution of relativistic energy inside cosmic structures. 

\subsubsection{Energy dissipation}
\begin{figure}
\includegraphics[clip,trim={0.3cm 0.3cm 0.3cm 0}]{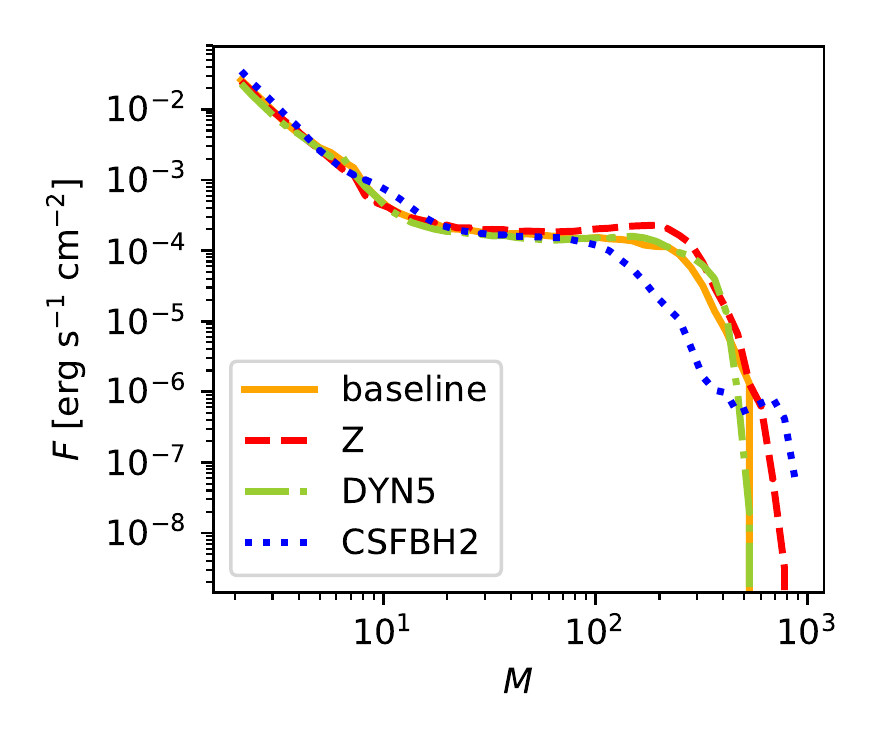}
\caption{Total kinetic flux per intervals of Mach number for the four Chronos simulations. The most energetic shocks ($M<2$) have been excluded, as explained in \ref{sec:chpr}.}
\label{fig:fkin}
\end{figure}

We define the incident kinetic energy flux as in \citet[][]{ry03}:
\begin{equation}
    F_{\mathrm{kin}}=\frac{1}{2}\rho_{\mathrm{pre}}v^3_{\mathrm{sh}}.
\end{equation}
 The total incident kinetic energy flux is represented in Figure \ref{fig:fkin}: the trend, already expected from Figure \ref{fig:mach}, reflects the dual distribution of shocks, i.e. internal (low $M$), with a monotonic decreasing flux, and external (high $M$), with a bump at $M\approx 100$. The behavior of CSFBH2 slightly deviates for $M \lesssim 5$ and $M \gtrsim 100$, for the 
same effects discussed in the previous Section.  
The dissipated incident kinetic energy flux which is converted into thermal energy flux $F_{\mathrm{th}}$, or into a CR energy flux $F_{\mathrm{CR}}$, can be parametrized as $\delta(M)=F_{\mathrm{th}}/F_{\mathrm{kin}}$ and $\eta(M)=F_{\mathrm{CR}}/F_{\mathrm{kin}}$, i.e. with the thermalization efficiency and with the assumed CR acceleration efficiency respectively. 
Although more recent works provided updated guesses for the acceleration efficiency by shocks, for the sake of comparison with previous works we based our  prescriptions for $\delta$ and $\eta$ on \citet[][]{kr13}, which assumed DSA as the accelerating mechanisms and included the effect of magnetic field amplification by CR streaming instabilities and Alfv\'enic drift.

\begin{figure*}
\includegraphics[clip,trim={0.1cm 0 0.1cm 8cm}]{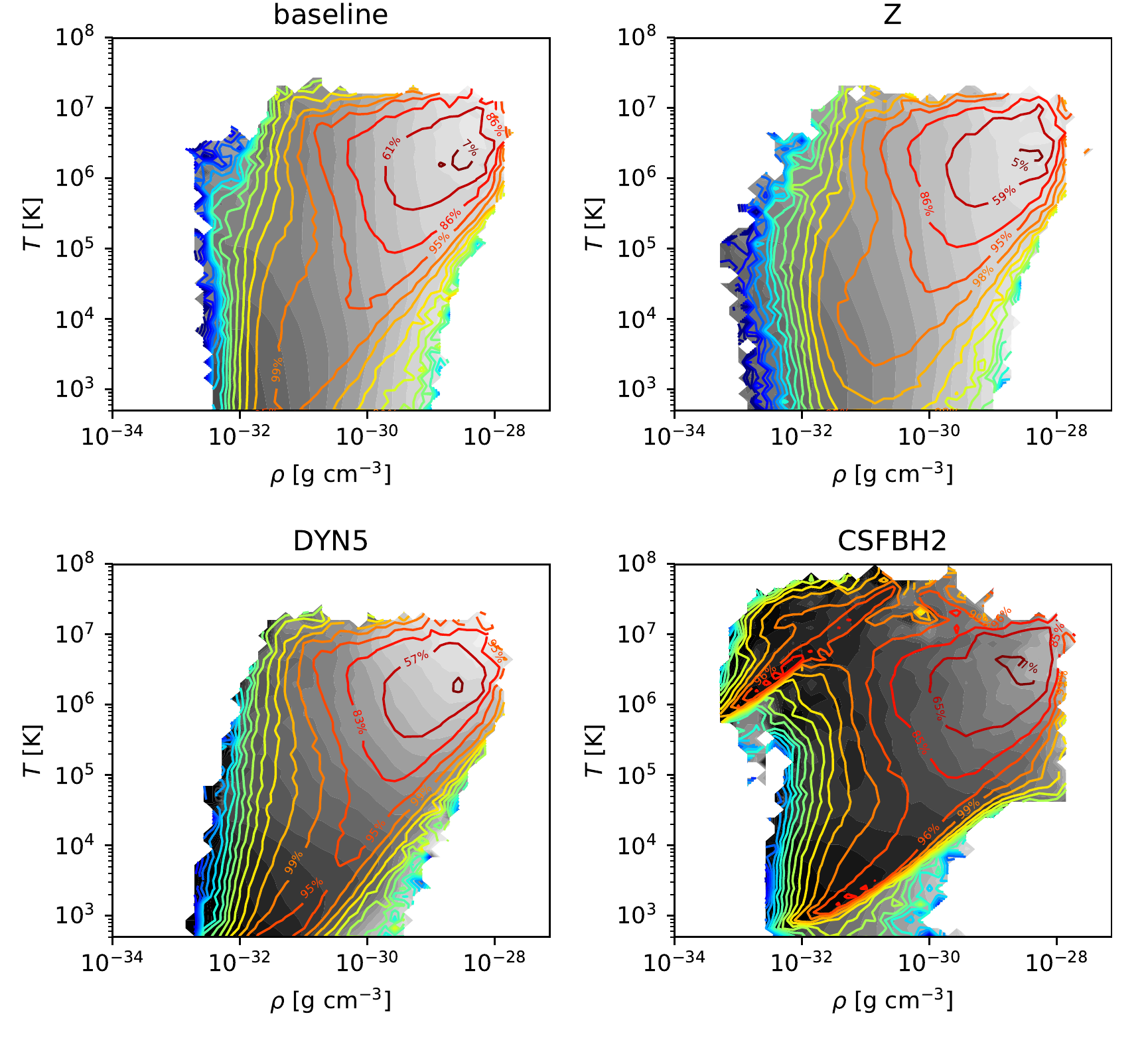}
\includegraphics[clip,trim={0.1cm 17cm 0.1cm 5cm}]{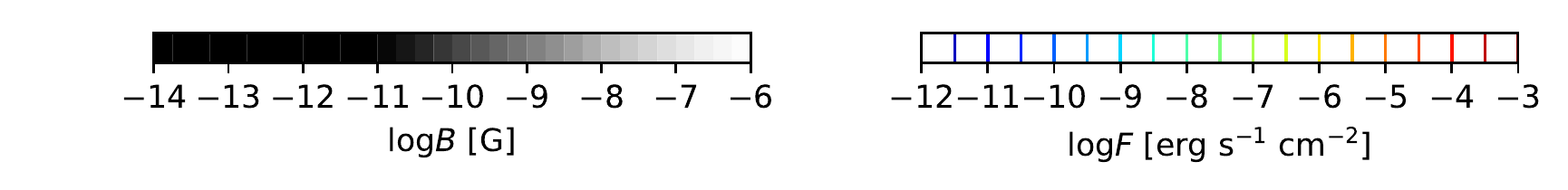}
\caption{Phase diagrams of shocked cells showing the pre-shock magnetic field in grey scale per intervals of pre-shock gas density and temperature for the four Chronos simulations at $z=0$. The colored contour lines show the sum of the dissipated flux in each bin, while the percentages indicate the fraction of the total flux which is dissipated for values of the flux higher than the corresponding one.}
\label{fig:phase}
\end{figure*}

The $(\rho,T)$ phase diagrams of the shocked cells indicating the values of pre-shock magnetic field and dissipated flux are given in Figure \ref{fig:phase}. In all the four runs, high-$M$ accretion shocks in low-density and low-temperature areas (see Figure \ref{fig:npp_d}) are less energetic (as in Figure \ref{fig:fkin}) and occur in higher-$\beta$ plasma. Instead, the most energetically relevant events occur in denser and hotter environments, through low-$M$ internal shocks: $\sim 85\ \%$ of the total energy flux in the volume is enclosed in the area having approximately pre-shock values of $\rho\gtrsim 10^{-30}\ \mathrm{g\ cm^{-3}}$ and $T\gtrsim10^{5}\ \mathrm{K}$ in all runs. Therefore, without taking into account the effect of shock obliquity on the acceleration of CR,  we can expect the bulk of CR acceleration to happen in the same environment in all models, i.e. within and around galaxy clusters/groups of galaxies. 
On the other hand, the four simulations are characterized by very different pre-shock magnetic field strengths: as a consequence of DYN5 and CSFBH2 having very weak seed fields, shocks in rarefied regions run over magnetic field strengths which are several orders of magnitude below the ones in the baseline and Z run. We remark that the shape of the phase diagram in CSFBH2 differs from the others due to the inclusion of reionization, which increases the temperature of pre-shock regions at low cosmic densities: this delimits a region at the bottom of the diagram (below the sharp discontinuity marked by contours of magnetic field and energy flux) whose combinations of temperature and density are forbidden for the intracluster medium (ICM). Shocks below this line are likely to be related to outflows originated in dense regions, whose temperature has cooled down, while its associated magnetization has only be affected by adiabatic expanse \citep[e.g][]{va17cqg}. However, these cells only process a negligible fraction of the total flux.
In summary, despite the very dissimilar magnetic properties of the simulations, we found a consistent trend as regards the thermal characterization of shocks and the energy dissipation.

\begin{figure}
\includegraphics[clip,trim={0.2cm 0cm 0.3cm 0cm}]{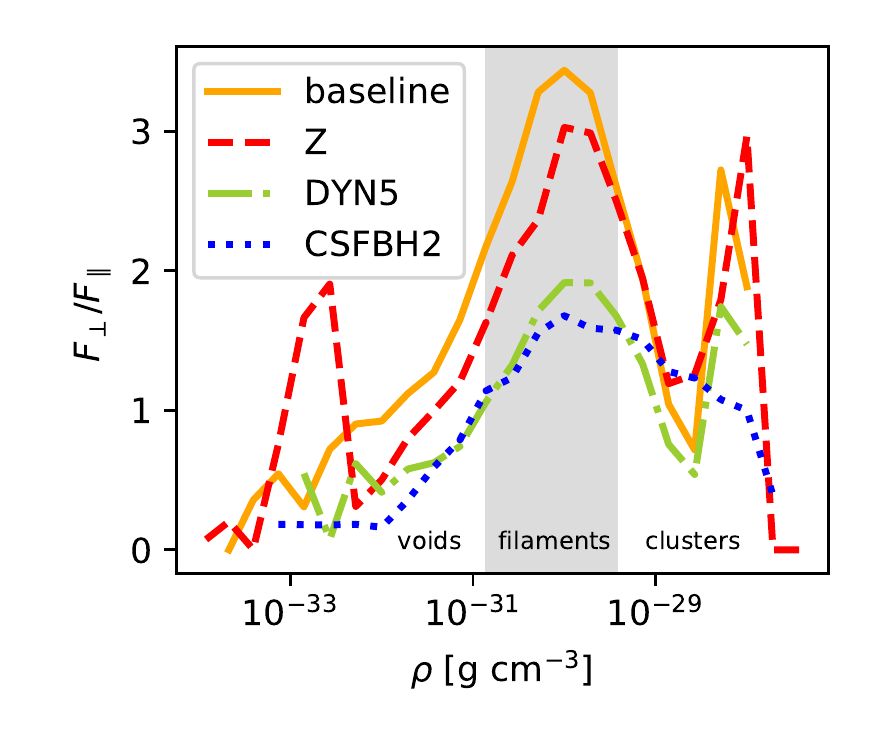}
\caption{Ratio of CR flux in quasi-perpendicular shocks and CR flux in quasi-parallel shocks as a  function of gas density for the four Chronos simulations at $z=0$.}
\label{fig:fpp}
\end{figure}

Figure \ref{fig:fpp} shows the ratio of the total CR energy flux in quasi-perpendicular shocks over the total CR energy flux in quasi-parallel shocks within gas density bins. While in the main paper we adopt the simple approach of setting the flux dissipated by quasi-perpendicular shocks to $0$ if $\Theta < 45^{\circ}$, or vice versa for quasi-parallel shocks, in Appendix \ref{app:parper} we present tests using smoother transition of efficiencies, which suggest similar outcomes.
While the global trends are in line with Figure \ref{fig:npp}, relating the $F_{\perp}/F_{\parallel}$ to the random distribution is here made difficult by the weighting for the energy flux, which can span across $\sim 10$ orders of magnitude in most environment, as shown above (Figure \ref{fig:phase}). As a consequence of this, a few energetic events can introduce large spikes in Figure \ref{fig:fpp}, which makes it harder to compare this to the expectation from random models. In general, the predominance of quasi-perpendicular shocks across environment and the relative differences between models are the same already discussed in the previous Section. %

\subsubsection{Galaxy cluster properties}

We identified galaxy clusters in the Chronos simulations using a standard algorithm, which delimits the virial volume of simulated halos based on the total matter overdensity, averaged assuming spherical symmetry with respect to the cluster center of mass \citep[e.g.][]{gpm98b}.  We performed the following analysis on the ten most massive clusters, which are in the mass range $5\cdot 10^{13}\ \mathrm{M_\odot}\lesssim M_{100}\lesssim 3 \cdot 10^{14}\ \mathrm{M_\odot}$\footnote{$M_{100}$ is defined as the total mass enclosed in a spherical volume of radius $r_{\mathrm{virial}}$, i.e. the distance from the cluster center where the average inner matter density is $100$ times the cosmological critical density.}. 

\begin{figure}
\includegraphics[clip,trim={0.2cm 0cm 0.3cm 0cm}]{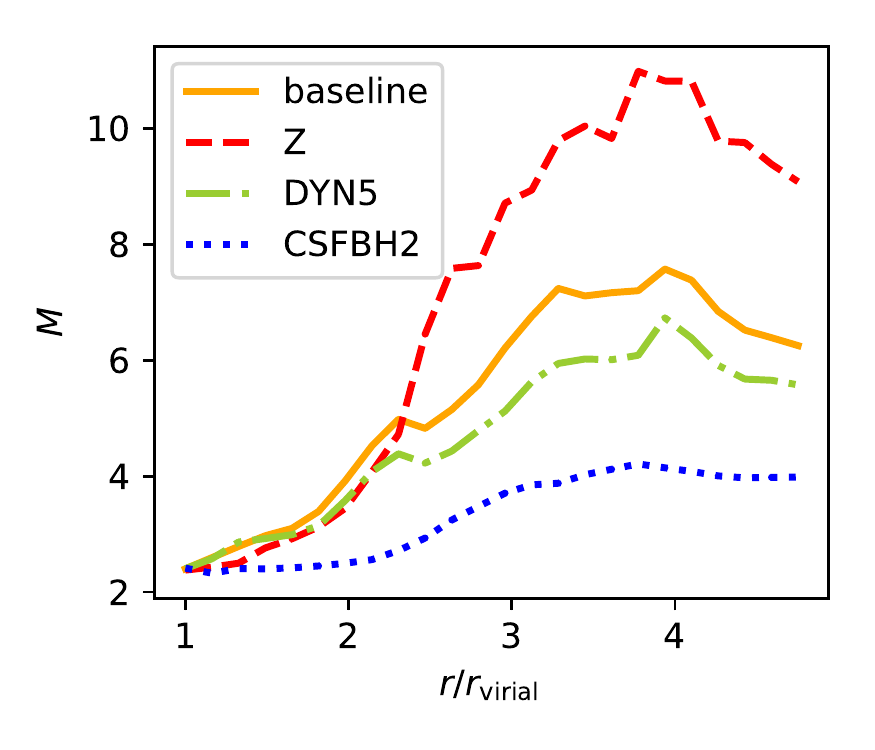}
\caption{Median Mach number as a function of radial distance from the cluster cores in units of virial radius in the range $1<r/r_{\mathrm{virial}}<5$. Values of $M$ span from $\approx 2$ to $\approx 10$: the CSFBH2 run hosts weaker shocks due to the higher temperatures.}
\label{fig:rm}
\end{figure}

\begin{figure}
\includegraphics[clip,trim={0.2cm 0cm 0.3cm 0cm}]{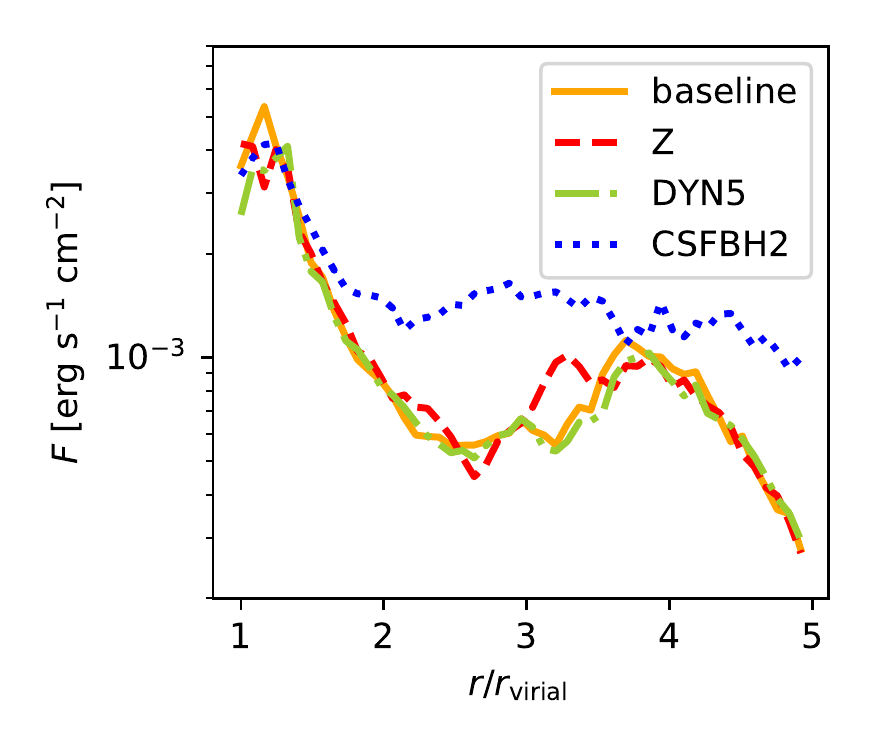}
\caption{Total dissipated kinetic flux as a function of the distance from the centers of the clusters, rescaled to the virial radius of each cluster, for the four Chronos simulations. The most energetic shocks ($M<2$) have been excluded.}
\label{fig:fprof}
\end{figure}

\begin{figure}
\includegraphics[clip,trim={0.2cm 0cm 0.3cm 0cm}]{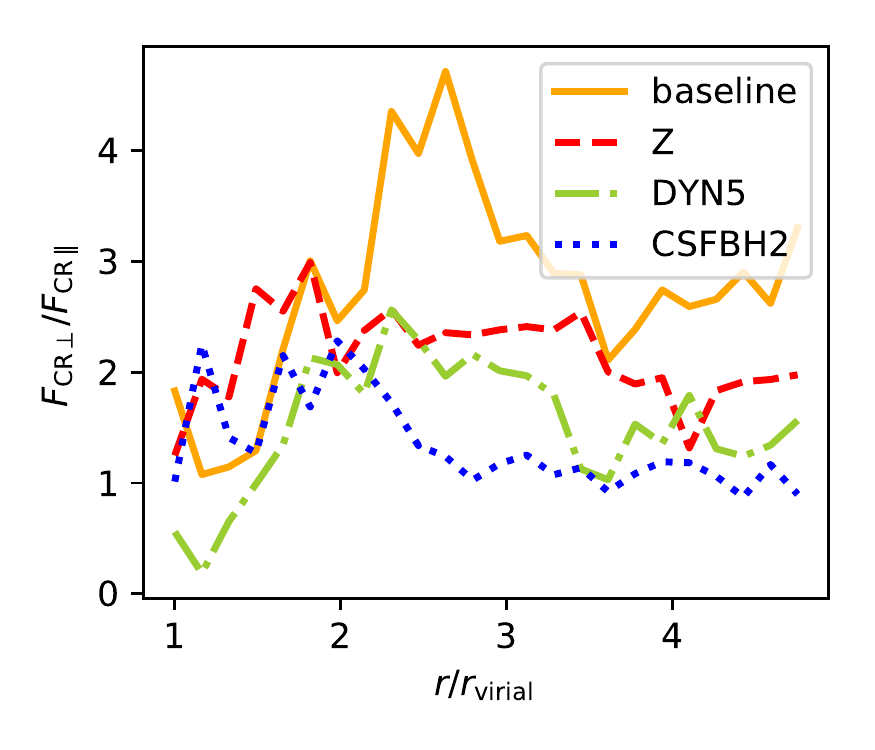}
\caption{Ratio of CR flux in quasi-perpendicular shocks and CR flux in quasi-parallel shocks as a function of the distance from the center of the clusters, rescaled to the virial radius of each cluster, for the four Chronos simulations.}
\label{fig:fcrprof}
\end{figure}

We extracted the radial profile of shocked cells inside these clusters for each of the four runs at redshift $z=0$. In Figure \ref{fig:rm}, we give the median value of shock Mach number as a function of the distance from the cluster, which shows an overall monotonic trend for all runs, in which the median Mach number of shocks increases as the local sound speed decreases following the radial decrease of gas temperature. The combination of the shallower radial trend of gas temperature due to reionization, as well as the integrated effect of shocks previously launched by the past activity of AGN explains the flatter radial behaviour of the Mach number profiles of clusters in the CSFBH2 model. 
Figure \ref{fig:fprof} shows the flux dissipated by shocks for each radial bin: shocks in CSFBH2 are globally more energetic in the proximity of clusters due to the additional driving of powerful but low-$M$ shocks promoted  by AGNs. Even higher fluxes are expected for shocks more internal than $1\ r_{\mathrm{virial}}$, but the averaging procedure blurs them in the simulated dataset, due to limited resolution and to the $M=2$ lower threshold (see Section \ref{sec:res_sp} for a more resolved view using nested grids). 
Finally, in Figure \ref{fig:fcrprof} we measure the ratio between the CR energy flux in quasi-perpendicular shocks over the CR energy flux in quasi-parallel shocks, as a function of radius from the cluster centers for the four runs. The trend of this ratio as a function of radius is similar to the one obtained as a function of gas density in Figure \ref{fig:fpp}, and further confirms the general trend that models with a large primordial seed field have a significantly larger dissipation of shock energy flux through quasi-perpendicular shocks, even at distances of $\sim 3-4\ r_{\mathrm{virial}}$ from the center of clusters.

We can recap the main results achieved by the analysis of the low-resolution cosmic volumes by saying that an excess of quasi-perpendicular shocks is widespread in all magnetization scenarios and in most cosmic environments. The extent of this tendency to produce perpendicular shocks has proven to be a function of magnetic properties, gas density, shock strength and distance from the clusters.  In particular, we consistently report that in shocks surrounding filaments the excess of quasi-perpendicular shocks is extreme, rather independently on the assumed magnetization scenario. 

In order to better assess the significance of such results at higher resolution, as well as to resolve in time the process which brings simulated fields to align with filaments, in the next Section we apply the same methods to study higher resolution simulations of galaxy clusters.

\subsection{Temporal and spatially resolved analysis of galaxy clusters: the origin of the excess of quasi-perpendicular shocks}
\label{sec:res_sp}

Shocks in the volume of clusters from the San Pedro runs were identified using the same procedure as for Chronos, except for the Mach number threshold: there, we neglected shocks below $M=2$ for the larger simulations, since they are not expected to be able to accelerate particles \citep{Ha2018protons}. On the other hand, with  San Pedro simulations we aim at studying the formation of perpendicular and parallel shocks, regardless of their strength and possibly even in the innermost hot regions of galaxy clusters, so in this case we set the threshold at $M=1.3$, which gives us a slightly higher statistics of shocks.  Probing weaker Mach numbers gets also more accurate at high resolution, while on coarser grids spurious classification can occur.

\begin{figure}
\includegraphics[clip,trim={0.2cm 0cm 0.3cm 0cm}]{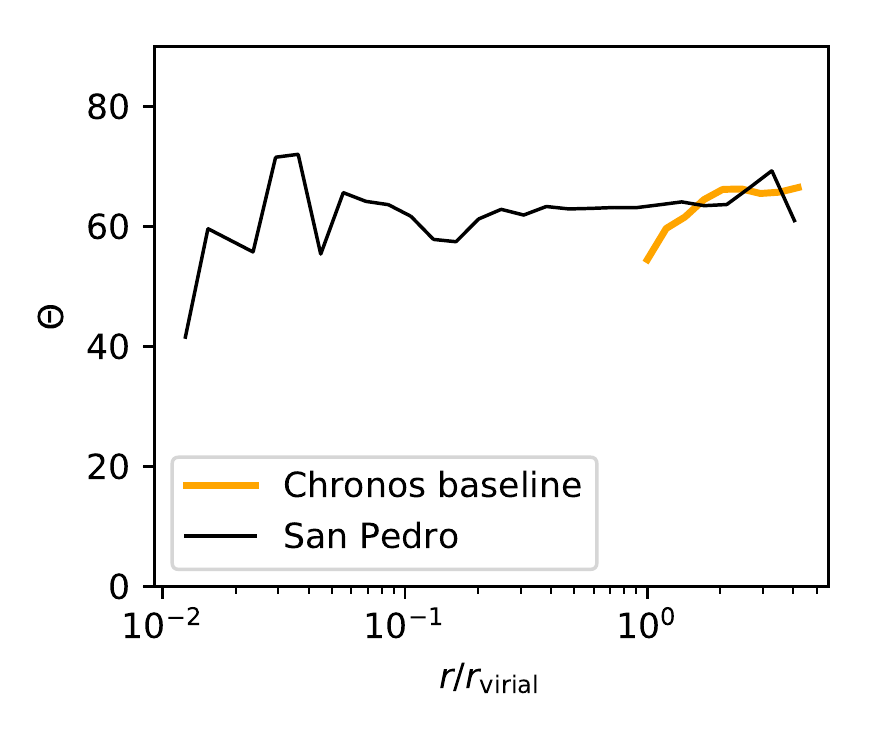}
\caption{Median obliquity as a function of the distance from the centers of the clusters, rescaled to the virial radius of the six San Pedro clusters, compared to the ten clusters extracted from the baseline Chronos run.}
\label{fig:spprof}
\end{figure}

\begin{figure*}
\includegraphics[scale=1,clip,trim={0.5cm 0.2cm 0.5cm 0.1cm}]{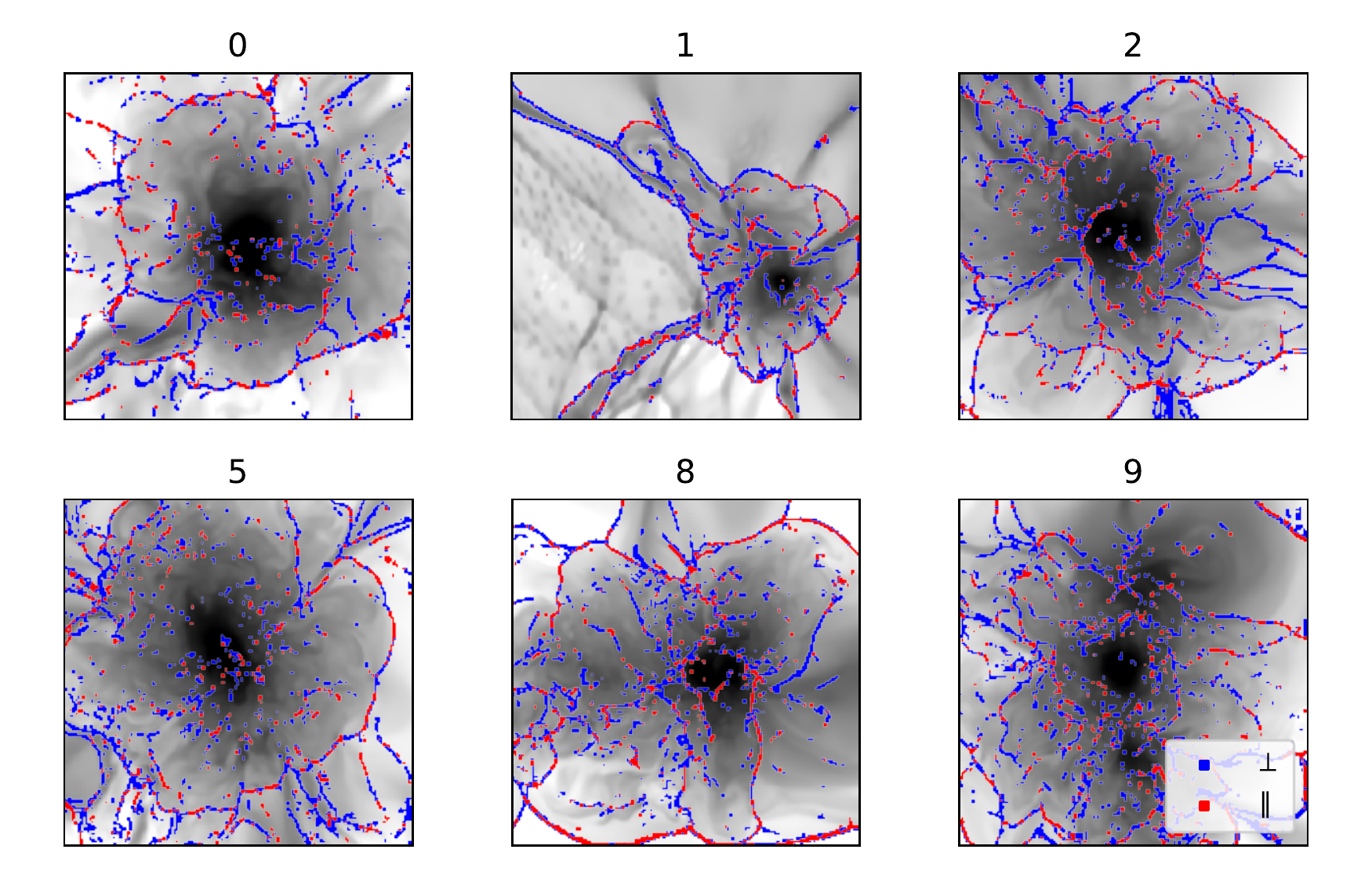}
\caption{Central slice of the San Pedro galaxy clusters. The grey scale measures the gas density, shocked cells are identified in blue (quasi-perpendicular) and red (quasi-parallel). The box length varies from $\approx 6.5\ \mathrm{Mpc}$ to $\approx 9.8\ \mathrm{Mpc}$.}
\label{fig:sp}
\end{figure*}

First, we assess whether there is a regularity between the two sets of simulations, in particular between the San Pedro clusters and the baseline simulation from Chronos, whose initial conditions are similar, and we investigate the reason behind the quasi-perpendicular excess found in Chronos. Figure \ref{fig:spprof} shows the median obliquity of shocks for each bin of radial distance from the cluster centers: we find there is an agreement above $1 \ r_{\mathrm{virial}}$, while there are too few identified shocks closer to the cluster cores in Chronos, due both to the lower resolution and to the higher Mach number threshold.
The distribution of shocks in the central slice of the high-resolution clusters can be seen in Figure \ref{fig:sp}, along with their color-coded obliquity.

\begin{figure*}
\includegraphics[scale=1,clip,trim={0.2cm 0.2cm 0.2cm 0.2cm}]{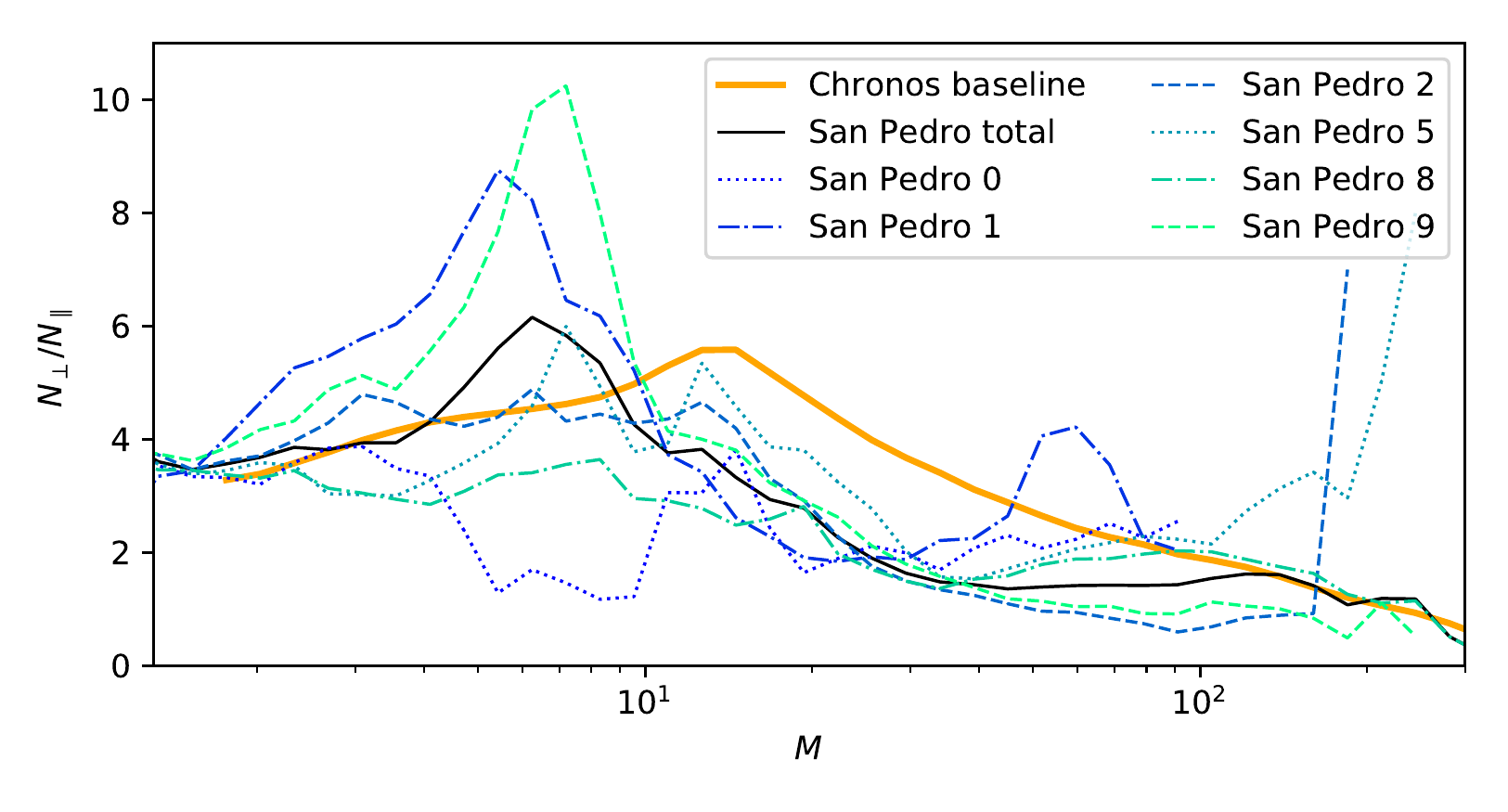}
\caption{Ratio of quasi-perpendicular to quasi-parallel shocks per Mach number intervals for the six San Pedro clusters (both individually and collectively), compared to the baseline Chronos run.}
\label{fig:parpersp}
\end{figure*}

The high-resolution clusters show an excess of quasi-perpendicular shocks, which is visually rendered by the predominance of blue cells in Figure \ref{fig:sp}. 
Figure \ref{fig:parpersp} shows the excess of quasi-perpendicular to quasi-parallel shocks as a function of Mach number for the six clusters, which is mostly prominent in the $5\lesssim M \lesssim20$ range, similar to the previous statistics derived for the Chronos suite. The shift in the peak from Chronos to San Pedro simulations is due to the $\sim 3$ times better resolution of San Pedro runs, which affects the Mach number estimation (see Section \ref{sec:shfin}).

Shocks propagating from the clusters outwards are mostly parallel, while quasi-perpendicular shocks are often associated to filaments (Figure \ref{fig:sp}): thanks to the higher resolution we now have a clearer view of what happens in the pre-shock region outside filaments. Due to the shear of the velocity field where filaments form starting from the cosmological initial conditions, derived from the Zeldovich approximation \citep{1996Natur.380..603B}, the magnetic field lines are dragged by the gas and tend to align with the leading axis of filaments.

Such large-scale motions, described by the shear velocity tensor,
are primordial, in the sense that they are already contained in the initial conditions of cosmological simulations \citep[e.g.][]{Libeskind,Zhu_2017}. This leads to the presence of significant alignment of the velocity field and of filaments already very early ($z \gtrsim 20$) as well as 
to a persistence of the leading orientation of the shear tensor for large,  $\gtrsim 10\ \mathrm{Mpc}$ scales. These scales are manifestly larger than the ones involved in the formation of accretion shocks, as well as in the injection of vorticity within filaments \citep[e.g.][]{ry08}. Simulations have also shown that on such linear scales the dynamics of gas fully follows the one of dark matter \citep[e.g.][]{Zhu_2017}. As a consequence, both pre-shock and post-shock velocity fields are affected by this phenomenon, as well as magnetic field lines, which mostly passively follow the gas velocity due to the large kinematic plasma $\beta$ in this environment (see Section \ref{sec:chpr}).

\begin{figure*}
\includegraphics[scale=1,clip,trim={0.5cm 0cm 0.5cm 0cm}]{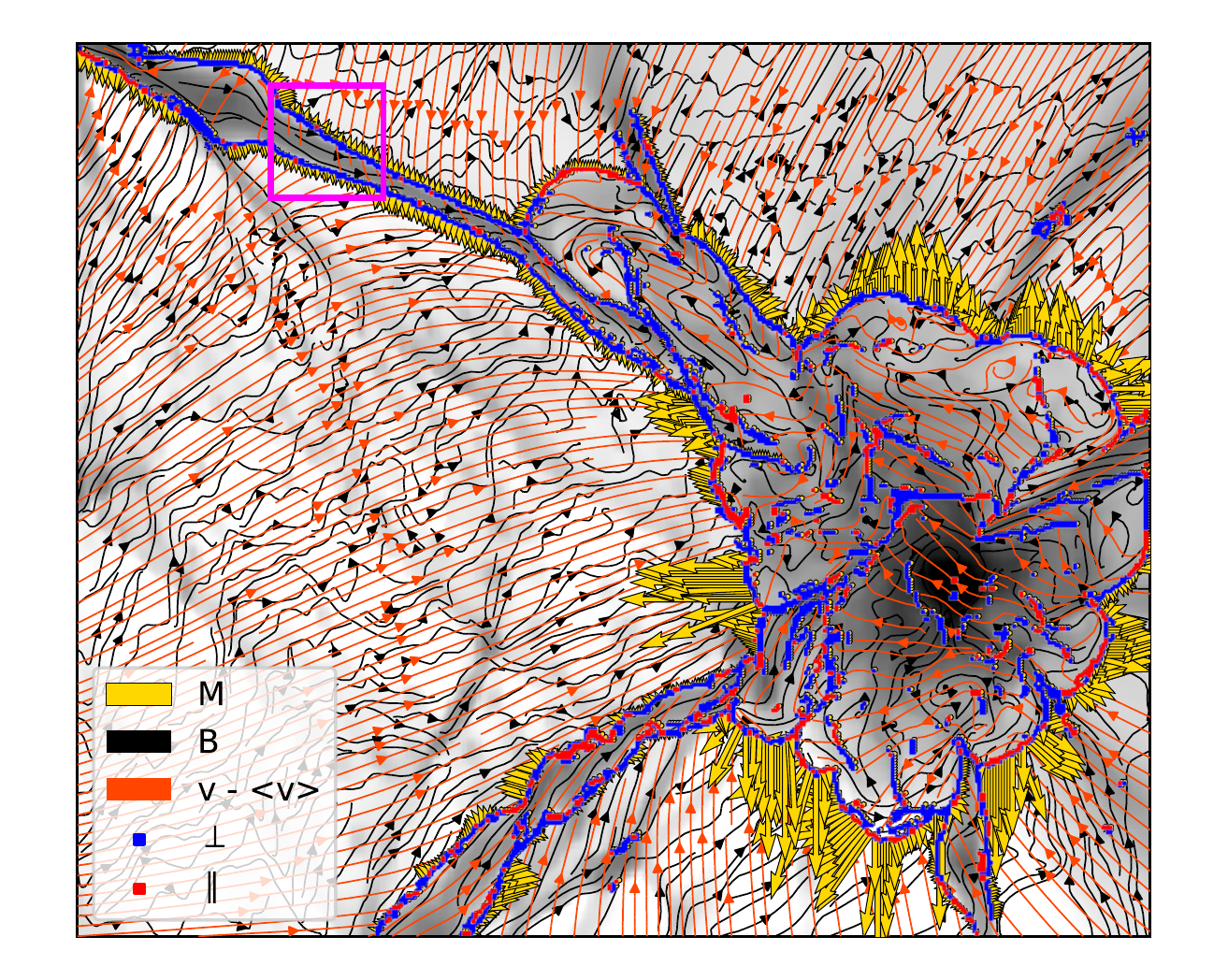}
\includegraphics[scale=1,clip,trim={4.2cm 0 4.2cm 0}]{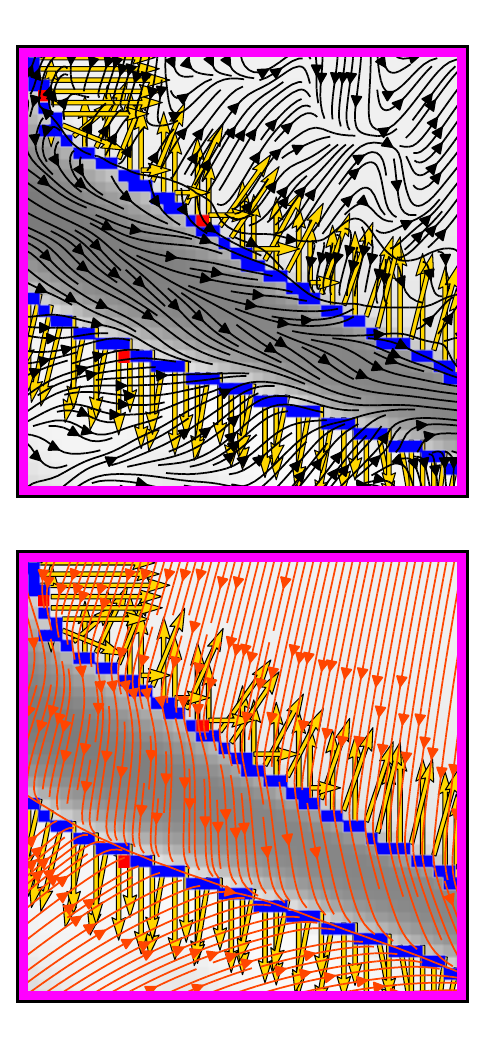}
\caption{Left panel: slice of one of the San Pedro galaxy clusters (number 1). The grey scale measures the gas density, the black arrows trace the projected magnetic field, the orange arrows trace the velocity field. Blue and red cells indicate shocks (respectively quasi-perpendicular and quasi-parallel shocks), the yellow arrows indicate the projected direction of propagation and intensity of the shocks. The box size is $\approx (9.6 \times 8.0)\ \mathrm{Mpc}^2$.  Right panel: zoom on a filaments from the left panel (box length of $\approx 1\ \mathrm{Mpc}$), where the bending of the magnetic field (top) and velocity field (bottom) lines along the filament in the pre-shock region can be seen.}
\label{fig:sp1}
\end{figure*}

\begin{figure*}
\includegraphics[scale=1,clip,trim={0.3cm 0.3cm 0.3cm 0.3cm}]{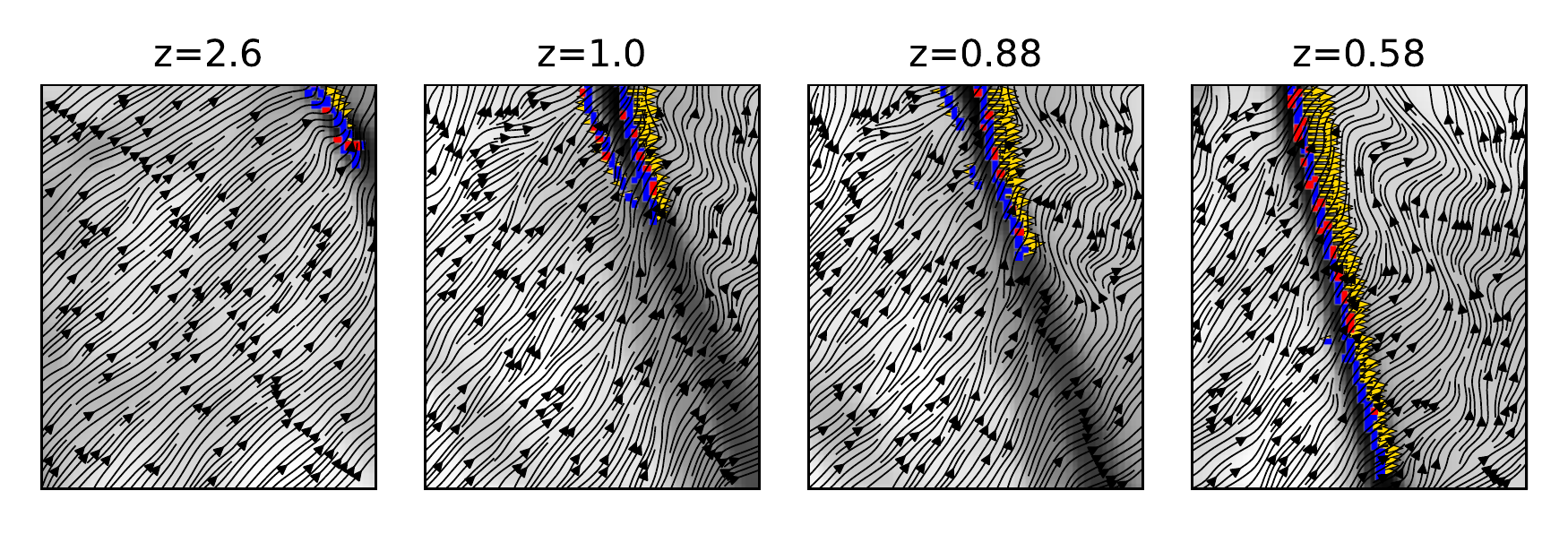}
\caption{Time evolution of a filament near San Pedro cluster number 1, where the represented quantities are the same as Figure \ref{fig:sp1}: magnetic field lines start to bend outside the filament before shocks are even generated, thus favoring perpendicular obliquities. The box size is $\approx (1.5 \times 1.8)\ \mathrm{Mpc}^2$.}
\label{fig:sp1ev}
\end{figure*}

The effect of shear motions which develop in the formation region of filaments can be seen in Figure \ref{fig:sp1}, where the streamlines of both velocity\footnote{The considered quantity is actually $\mathbfit{v}-\langle\mathbfit{v}\rangle$, where $\langle\mathbfit{v}\rangle$ is the velocity field averaged inside the whole slice. Without this adjustment, the flow motions would be dominated by a bulk displacement towards more massive centers of gravity lying outside the box and the smaller-scale flow around filaments would not be visible.} and magnetic field tend to self-align with the leading axis of filaments. As this large-scale velocity shear emerges from the cosmological initial conditions and is structured on scales larger than the filament width, the local alignment affects both the pre-shock and the post-shock regions across the filament's edge. This can be observed in Figure \ref{fig:sp1ev}, where the time evolution of the arrangement of the magnetic field around a filament is shown. Therefore, when shocks are formed at the interface between the infalling smooth gas accreted from voids and the filament's region, they often propagate over a magnetic field which was previously aligned with the filament axis by the large-scale shear. This leads to the tendency of forming mostly quasi-perpendicular shocks in the simulated cosmic web. 
The framework recently developed by  \citet[][]{Soler2017} (and applied to study the effects of MHD turbulence on the distribution of angles formed by density gradients, $\nabla \rho$, and magnetic fields in simulations of the interstellar medium) provides a quantitative explanations for this effect. 
They found that the alignment of the magnetic field along the direction of low-density structures (e.g. filaments) is spontaneously produced in regions where shear motions dominate over compression. In particular, they demonstrated that  the configuration where $\nabla \rho$ and $\mathbfit{B}$ are mostly parallel or mostly perpendicular are equilibrium points to which the fluid 
tends to evolve on hydrodynamical timescales, depending on the local flow conditions. Supported by direct numerical simulations, they reported  that for low density, high $\beta$ and large (negative) velocity divergence the relative orientation of magnetic fields and $\nabla \rho$ becomes preferentially quasi-perpendicular, and this well explains what we also observe in our simulations (see also Appendix \ref{app:xi}). Considering that $\nabla \rho$ is in general a trustworthy tracer of the shock direction, this explains why shocks in our simulated filaments have the tendency of showing quasi-perpendicular geometries. 

As a consequence, the excess of perpendicular shocks over parallel shocks (Figure \ref{fig:parpersp}) varies from cluster to cluster due to their different evolutionary stages, as well as depending on the number of filaments connected to them. 


\begin{figure}
\includegraphics[scale=1,clip,trim={0.4cm 0.5cm 0.4cm 0.2cm}]{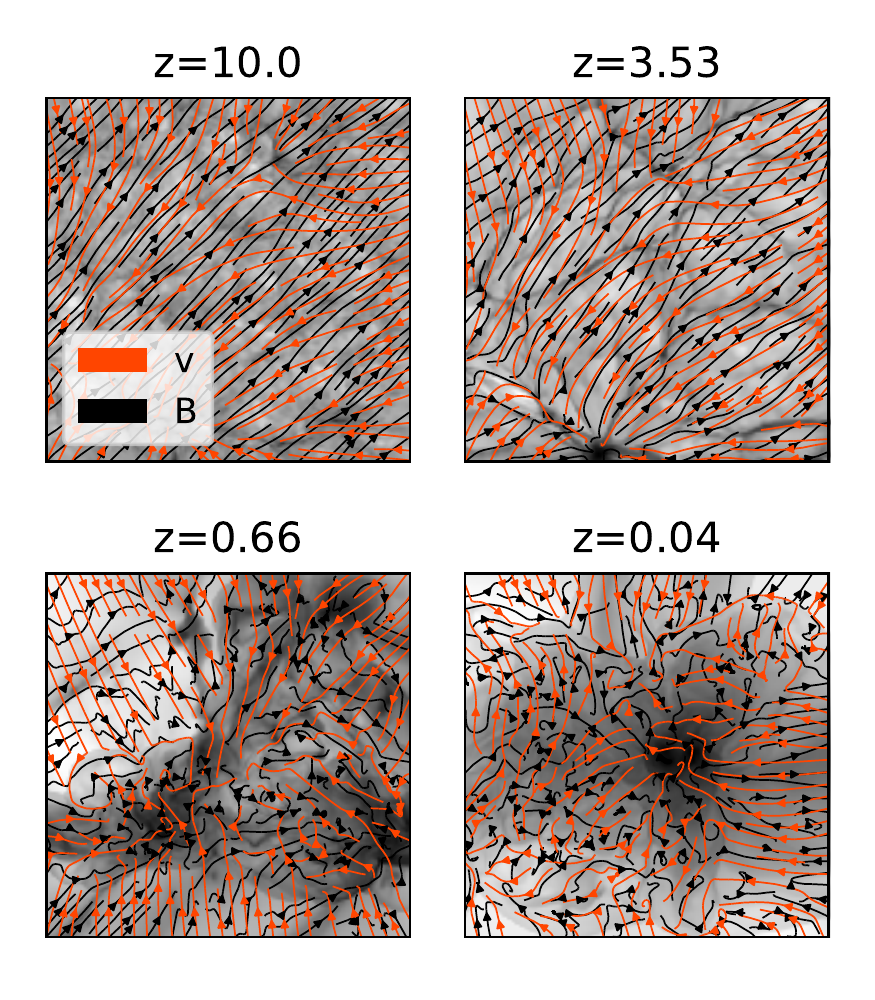} 
\caption{Central slice of one of the San Pedro galaxy clusters (number 8) at different redshifts $z$. The grey scale measures the gas density, the black arrows trace the projected magnetic field and the orange arrows trace the projected velocity field (box length of $\approx 9.6\ \mathrm{Mpc}$).}
\label{fig:spevol}
\end{figure}

\begin{figure}
\includegraphics[scale=1,clip,trim={0.2cm 0.5cm 0.3cm 0.1cm}]{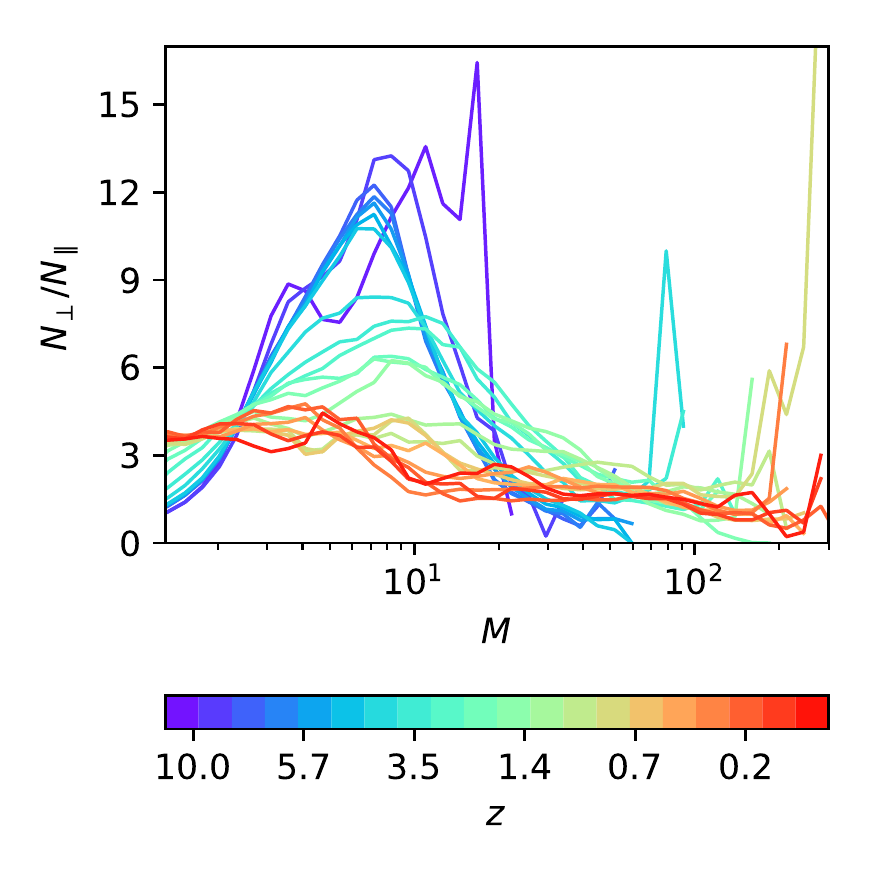}
\caption{Time evolution of the ratio of quasi-perpendicular to quasi-parallel shocks per Mach number intervals for San Pedro cluster number 8 from redshift $z=10$ to $z=0.1$.}
\label{fig:parperz}
\end{figure}

Finally, Figure \ref{fig:spevol} shows the evolution from $z=10$ to $z=0.1$ of one of the clusters with streamlines of magnetic field and velocity: at the beginning the orientation is set by the initial conditions, but later on the dynamics begin to dominate and the fields self-arrange in a pattern that favors perpendicular shocks near filaments. The evolution of the quantity $N_{\mathrm{\perp}}/N_{\mathrm{\parallel}}$ is shown in Figure \ref{fig:parperz}: higher values of this volume-weighted statistics are typically reached at high redshifts, given that the box includes more filaments before the cluster has grown to its final volume. As $z$ decreases, the excess of perpendicular shocks decreases because the same volume gets increasingly more swept by internal merger shocks running over a typically tangled magnetic field. 

In summary, our analysis supports that the circulation of gas falling into filaments, coupled via MHD equations to the evolution of magnetic fields \citep[][]{Soler2017}, is responsible for the observed tendency of shocks in filaments to be preferentially perpendicular to the local orientation of the magnetic fields. In retrospective, this model can also explain why the same tendency is somewhat less prominent in the CSFBH2 and in the DYN5 models, as previously outlined in Section \ref{res_chronos}. In the latter models, the magnetic field is subject to a more significant build-up over time, due to either the effect of the injection of new magnetic fields via feedback events, or due to the implemented sub-grid dynamo amplification. As a result of this, in these models shocks are typically running over dynamically ``young'' magnetic structures, in the sense that the alignment mechanisms described by \citet{Soler2017} is less effective there, because the equilibrium point in the local topology of magnetic fields and density gradient can only be reached after a fraction of the system crossing time. Moreover, the impulsive activity by AGN feedback continuously stirs fluctuations in the surroundings, making it difficult for the fluid to equilibrate. For the above reasons, we can thus conclude that the excess of quasi-perpendicular shock geometries is a tendency found regardless of the model of magnetogenesis, and that the amplitude of this excess is increased in primordial models, or in general in scenarios where magnetic fields have co-evolved with gas matters for longer timescales. 


\section{Implications for observations}
\label{sec:obs}
The distribution of shock obliquities in the cosmic environment and in galaxy clusters may have a significant impact on the observed CR signatures, i.e. synchrontron radio emission produced by CR electrons and hadronic $\gamma$ rays from CR proton interactions. We briefly discuss that the estimates of obliquities found in our work suggest that $\gamma$-ray emission may be lower than expected and thus explain the \textit{Fermi}-LAT non-detections.

\begin{figure}
\includegraphics[scale=1,clip,trim={0.3cm 0.5cm 0.3cm 0}]{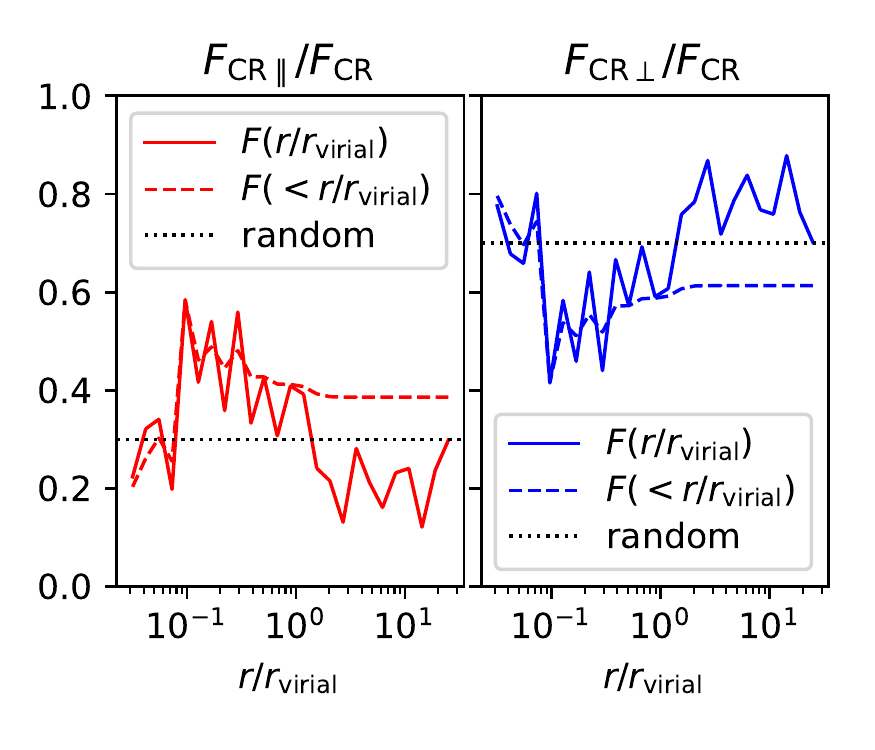}
\caption{Fraction of CR flux dissipated by quasi-parallel (left) and quasi-perpendicular (right) shocks as a function of distance from the San Pedro clusters cores. The solid line represents the total flux dissipated by shocks at a certain $r/r_{\mathrm{virial}}$, while the dashed line represents the integral of the same quantity enclosed inside a certain radius. The dotted line is the trend we would expect if obliquities were randomly distributed.}
\label{fig:gammarad}
\end{figure}

In Figure \ref{fig:gammarad}, we compute the differential and cumulative fraction of CR flux dissipated by parallel or perpendicular shocks with respect to the total $F_{\mathrm{CR}}$, as a function of distance from the San Pedro clusters' cores. For the sake of simplicity, we consider here that the impact of obliquity on the acceleration of radio-emitting electrons can be studied with respect to the instantaneous energy dissipation at shocks, while the impact on CR protons is more related to the total energy flux processed by shocks within the cluster volume.
Indeed, the characteristic lifetime of the synchrotron emitting electrons ($\sim\mathrm{GeV}$ energies) due to energy losses is $\lesssim 10^8\ \mathrm{yr}$ in clusters \citep{2019SSRv..215...16V}, which corresponds to a diffusion length-scale in the ICM of $10\ \mathrm{kpc}$ \citep{2002NewA....7..249B}. This implies that $F_{\mathrm{CR\perp}}(r)$, i.e. 
the sum of the CR energy dissipated by quasi-perpendicular shocks within each radial shell, directly relates to the energy dissipation involved in the powering of observable radio relics.  On the other hand, CR protons are long-lived  (i.e. typically longer than the age of the cluster for energies $\lesssim 2\cdot 10^7\ \mathrm{GeV}$, \citealt{bbp97}), and thus at a given epoch a better proxy for their level of hadronic $\gamma$-ray emission is given by the total integrated amount of CRs within the cluster radius.
Figure \ref{fig:gammarad} shows the average behavior of all shocks in the San Pedro sample, in which we binned the contribution from all clusters as a function of radius, normalized to the virial radius of each system. 
The plots show that $F_{\mathrm{CR\parallel}}/F_{\mathrm{CR}}$ lies mostly below $F_{\mathrm{CR\perp}}/F_{\mathrm{CR}}$, with a maximum difference at large radii from the cluster center, which based on Section \ref{sec:res_sp} is largely related to filaments. The total dissipation via quasi-parallel shocks is $\sim 40\ \%$ of the total CR flux within the virial radius, and only $\sim 20\ \%$ within the clusters core (dashed line, left panel).  The fraction of CR flux linked to perpendicular shocks is, on the other hand, larger than the random distribution at distances larger than the virial radius (solid line, right panel), and is $\sim 80\ \%$ of the total CR flux in most of the large volume surrounding the clusters virial radius.

\begin{figure*}
\includegraphics[scale=1,clip,trim={0.1cm 0.5cm 0.1cm 0}]{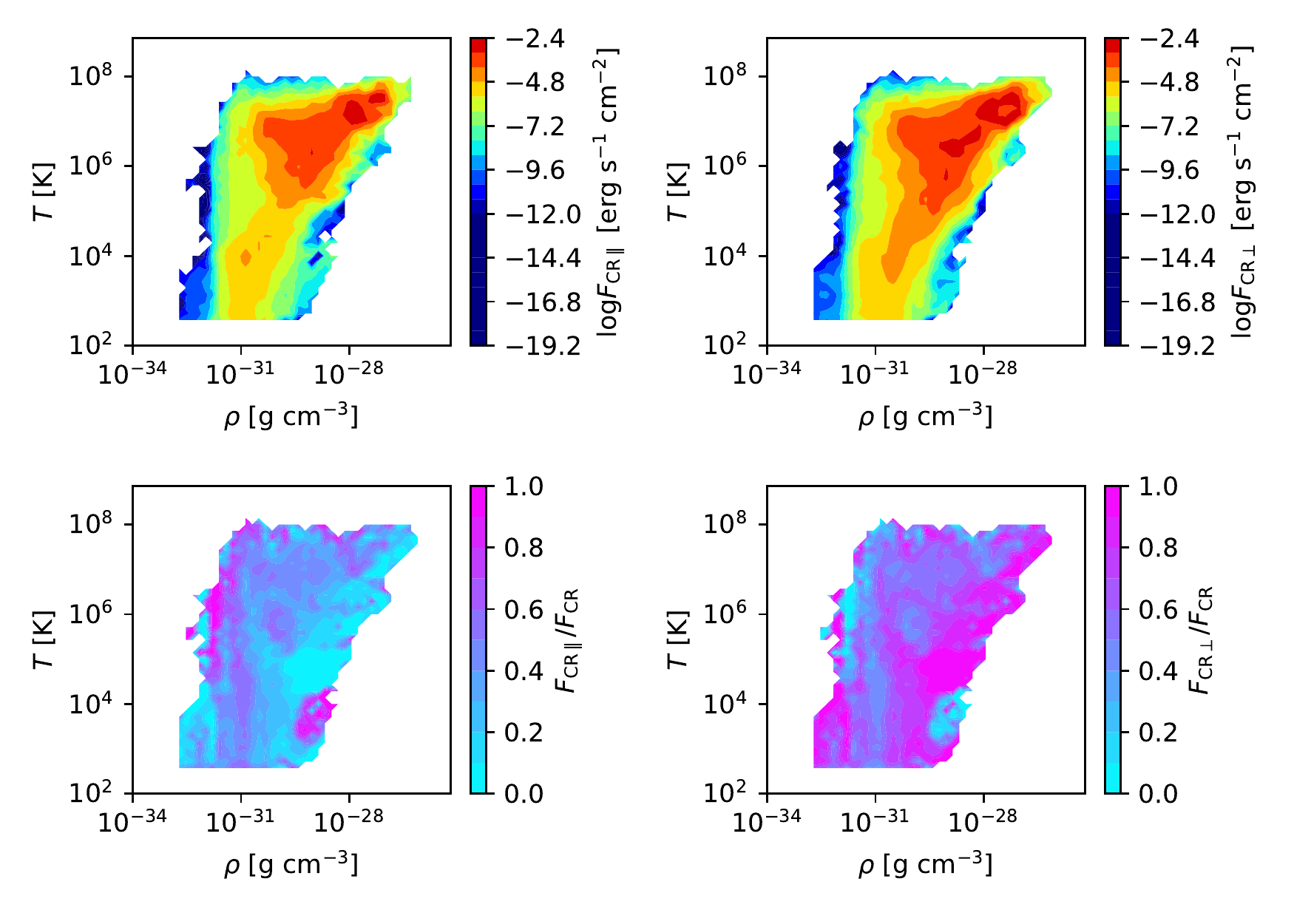}
\caption{Phase diagrams for the San Pedro clusters showing the CR flux dissipated by quasi-parallel and quasi-perpendicular shocks and their fraction with respect to the total as a function of pre-shock density and temperature.}
\label{fig:gammarad2}
\end{figure*}

Figure \ref{fig:gammarad2} shows the phase diagrams of the CR flux for quasi-parallel and quasi-perpendicular shocks in all San Pedro clusters. Similar to what found for the Chronos phase diagrams, the most energetic merger shocks are located 
in the temperature and density range typical of the innermost ICM, i.e. $T \gtrsim 10^{7}\ \mathrm{K}$ and $\rho \gtrsim 10^{-29}\ \mathrm{g\ cm^{-3}}$. Such shocks dissipate $\gtrsim 10^{-4}\ \mathrm{erg\ s^{-1}\ cm^{-2}}$ and should be responsible for most of the injection of CRs in the ICM, and are also likely connected with the powering of radio relics, \citep[e.g.][]{2019SSRv..215...16V}. From the phase diagrams, it can clearly be seen that most of their CR flux gets dissipated into quasi-perpendicular shocks, up to $\sim 70-80\ \%$. On the other hand, quasi parallel shocks are found to dominate the dissipation of CRs only for low density regions, which are however associated with negligible CR flux levels ($\lesssim 10^{-8}\ \mathrm{erg\ s^{-1}\ cm^{-2}}$). 

\begin{figure*}
\includegraphics[scale=1,clip,trim={0 4cm 1cm 0}]{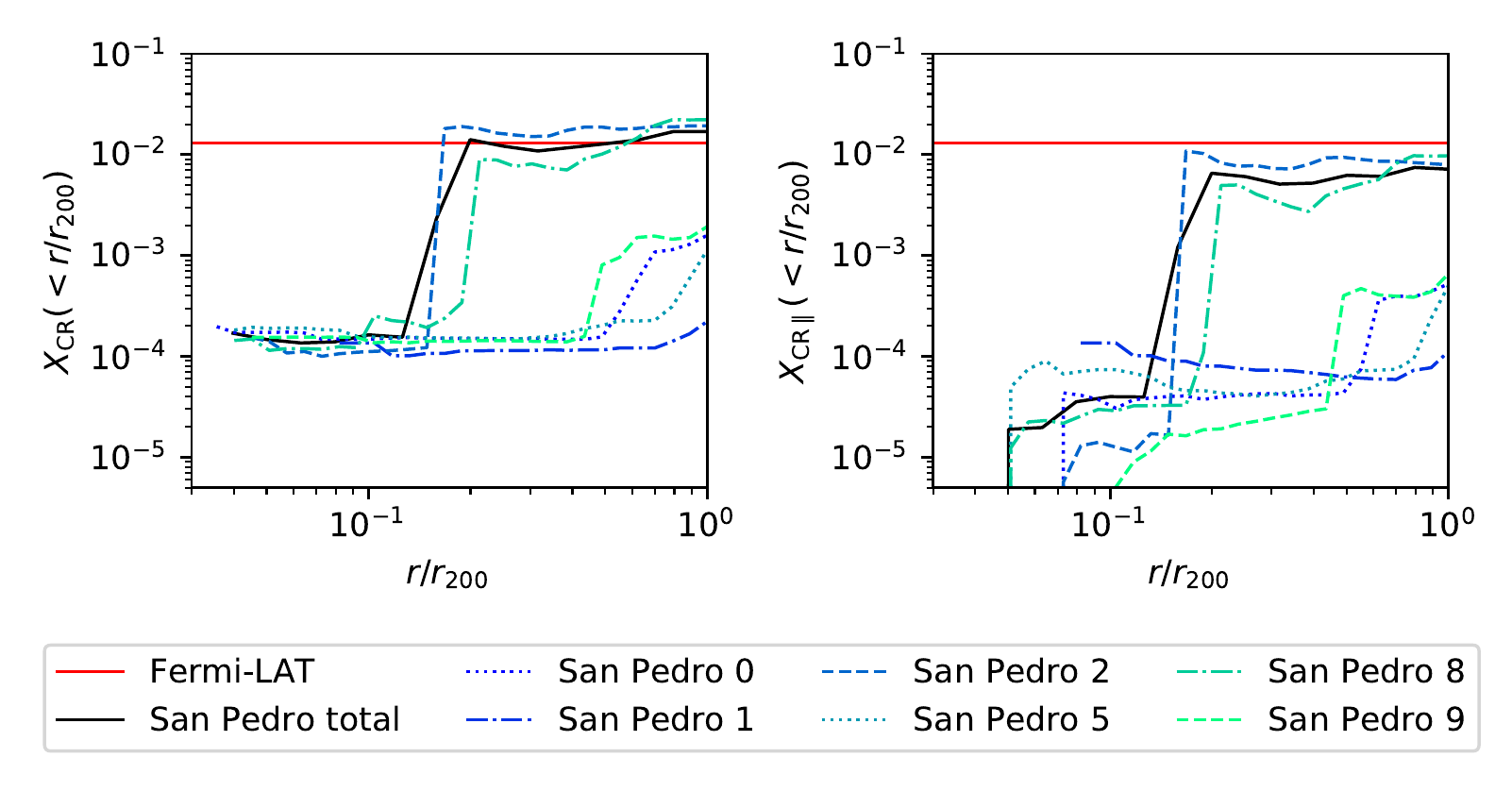}
\caption{CR pressure ratio for all shocks (left) and for parallel shocks only (right), integrated inside a certain radial distance from cluster cores (rescaled to $r_{200}$), compared to the upper limit imposed by \textit{Fermi} (red line) for the San Pedro clusters both globally (black solid curve) and individually (colored curves).}
\label{fig:gammarad3}
\end{figure*}

\textit{Fermi} observations \citep{fermi14} constrained the CR proton pressure ratio $X_{\mathrm{CR}}$, i.e. the ratio between the total CR proton pressure and the total gas pressure within $r_{200}$ for the observed cluster population at the level of $X_{\mathrm{CR}} \lesssim 1.3\ \%$. Under the assumption of a quasi-stationary population of shocks, and considering that the bulk of gas pressure and CR pressure is produced by shock dissipation, here for simplicity we can relate $X_{\mathrm{CR}}$ to the ratio of the instantaneous flux of CRs and thermal energy flux at shocks, i.e. $X_{\mathrm{CR}} \approx F_{\mathrm{CR}} / F_{\mathrm{th}}$, similar to other works \citep[e.g.][]{va09shocks,Ha2019}. 
We thus computed this ratio assuming that all shocks are able to accelerate protons, and compared this to the same quantity restricted to $< 45^{\circ}$. The two panels in Figure \ref{fig:gammarad3} compare the CR pressure ratio before and after the obliquity cut: if no distinctions were made in terms of obliquity, at least two of the simulated clusters in our sample would be detected by \textit{Fermi}, while all the curves of $X_{\mathrm{CR}}$ are shifted below the \textit{Fermi} detection threshold if only quasi-parallel shocks are considered. This suggests that our assumptions on obliquity could in principle be able to explain the missing detection of hadronic $\gamma$ rays, even with the CR acceleration efficiency by \citet{kr13} assumed here. 

\begin{figure}
\includegraphics[scale=1,clip,trim={0.3cm 0.5cm 0.3cm 0.5cm}]{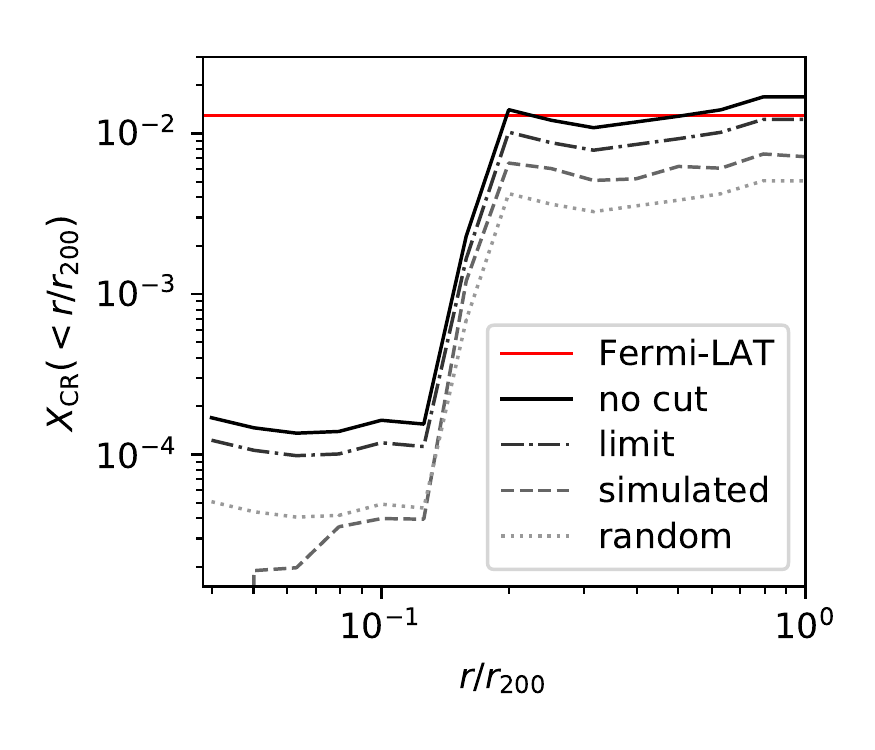}
\caption{CR pressure ratio of the totality of the San Pedro clusters for different obliquity distributions compared to the upper limit imposed by \textit{Fermi} (red line). The grey lines represent the CR pressure ratio for no obliquity cut (solid line), for the simulated obliquity (dashed line), if obliquity were randomly distributed (dotted line) and for the limit obliquity distribution above which $\gamma$ rays would be detected by \textit{Fermi} (dash-dotted line).}
\label{fig:gammarad4}
\end{figure}

In Figure \ref{fig:gammarad4} we show how the CR pressure ratio of the six clusters combined changes for different obliquity distributions. Although globally we obtained more perpendicular shocks than randomly expected, this is mostly relevant around filaments, while near cluster cores the trend is more similar to the random distribution of obliquities: as can be seen in Figure \ref{fig:gammarad} as well, shocks occurring inside the cluster virial radius can even have more parallel obliquities than random. As a consequence, the CR pressure ratio which would be obtained from a purely random distribution of shock obliquities (in which $\approx 30\ \%$ of shocks would have $\Theta \lesssim 45^{\circ}$) lies below the one found in the simulation (see dotted line in Figure \ref{fig:gammarad4}). We also estimated the upper limit of the fraction of shocks able to accelerate protons ($\approx 72\ \%$), such that a higher value would generate a marginally detectable $\gamma$-ray emission from our clusters within $r_{200}$ (dash-dotted line).

\color{black}

As a caveat, here we did not apply any temporal integration of the CR energy flux which might vary with time \citep[e.g. see Figure 7 in][]{wi17}. Furthermore, \citet{Ha2018protons} found that in a high-$\beta$ plasma the Mach number has to be larger than $2.25$ for efficient CR proton acceleration. However, although we do not consider a Mach number cut at $2.25$ that would further reduce the CR energy flux \citep[][]{Ha2019}, we adopted the values of $\eta$ from \citet{kr13}, which strongly decrease the contribution of low-$M$ shocks. As a consequence, the estimates of the CR pressure ratio would barely be affected by introducing this cut. Additionally, our findings are in agreement with a more sophisticated modelling of the $\gamma$-ray emission in the San Pedro simulations using Lagrangian tracer particles, recently discussed by \citet[][]{wittor20}.

\section{Numerical limitations and approximations}
\label{sec:disc}

We shortly address here (and more in detail in the Appendix) the  unavoidable numerical limitations involved in our analysis. 
First, the adopted spatial resolution somewhat affects the shock finding process, since pre-shock and post-shock conditions are often contaminated by additional flows which blend with the shock.
However, although there is a slight discrepancy between the Chronos and San Pedro simulations in the estimate of Mach numbers (e.g. Figure \ref{fig:parpersp}), we found an agreement for the trends of obliquity between the two sets of simulations (see Figure \ref{fig:spprof}).
The computation of obliquity itself is still a first attempt: the identification of the pre-shock cell and thus the reliability of the up-stream magnetic field orientation are highly affected by the resolution of the simulation (see Appendix \ref{app:parall}).
A close examination of the Chronos simulations exposed an additional issue, most likely of numerical origin, i.e. sequences of wave-like features in the magnetic field orientation in the most rarefied regions of the simulated volume. Low densities are usually critical to handle in simulations, as they are subject to hypersonic flows, for which a ``dual energy formalism'' is necessary \citep[e.g.][]{enzo14}, as well as to the introduction of numerical floor levels of gas temperature.  The fact that such patterns change their orientation when a different solver is used, unlike the rest of the simulation (see Appendix \ref{app:waves}), leads us to suspect that these features are driven by numerics and shall not be trusted. However, the impact of such biased obliquities is small in our statistics, as they are limited to very underdense regions and are associated to less than $2\ \%$  of the total energy dissipated by shocks in the volume. 

There are also physical processes that were not included in these simulations, which give this work room for improvement. In particular, no microphysical processes are implemented: according to DSA, strong shocks may undergo modifications by CRs, as well as magnetic field amplification promoted by CR-driven instabilities, turbulence generation and plasma heating \citep[for a review see][]{br11}.
Moreover, in this work (and similarly to previous others in the literature, e.g. \citealt{Ha2019}) we relied on estimates of the instantaneous energy dissipation into CRs to derive comparisons with the integrated energy budget constrained by \textit{Fermi}-LAT. More sophisticated approaches involving the deployment of ``two-fluid models'' \citep[e.g.][]{pf07,va16cr} or of passive tracer particles storing the evolution of CRs in time \citep[e.g.][]{wi17} were beyond the goal of this work, which is just the first step towards future ad-hoc PIC simulations.

\section{Summary and conclusions}
\label{sec:con}
In this work we have considered two sets of grid simulations performed with \textsc{Enzo}: Chronos, which simulates an $\approx (85\  \mathrm{Mpc})^3$ volume, and San Pedro, which zoomes in on six galaxy clusters, $\sim 3$ times more resolved than Chronos. We evaluated the typical obliquity of cosmic shocks as a function of environment, magnetic field topologies and magnetogenesis scenarios. We then used these results to estimate the observable CR flux for both proton and electron signatures and try to address the issue of the missing $\gamma$-ray detection by \textit{Fermi}-LAT.
The main conclusions we reached are the following:
\begin{enumerate}
    \item in the simulated cosmological volume, there is always an excess of quasi-perpendicular shocks over quasi-parallel shocks (up to $\sim 5$ times) for each of the Chronos runs. This excess is even more significant than the excess expected from a purely random distribution of angles in a three-dimensional space (Section \ref{sec:chpr}, Figure \ref{fig:npp});%
    \item this excess characterizes shocks generated in pre-shock regions with gas density of $\rho\approx 10^{-30}\ \mathrm{g\ cm^{-3}}$. The interval of Mach numbers with the highest percentage of perpendicular shocks is generally in the range $5 \lesssim M \lesssim 20 $, but it depends on the typical pre-shock temperatures, e.g. the runs with dynamo and reionization have higher $T$ and thus lower $M$ (Section \ref{sec:chpr}, Figure \ref{fig:npp_d});
    \item this effect is maximized around filaments, where the shocks propagate perpendicularly to the filament length and the magnetic field aligns with the filament (Figure \ref{fig:sp});
    \item the physical effect at play here likely is the progressive alignment of the magnetic field with respect to the velocity vector, through the equilibration process described by \citet{Soler2017}, which often arranges the local magnetic field to be quasi-perpendicular to the local direction of the density gradient;
    \item the alignment of the  magnetic field depends on the plasma $\beta$ and the magnetic field's origin, as well as its initial topology (Section \ref{sec:chpr}, Figure \ref{fig:npp}). A recent origin of magnetic field lines is found to slightly reduce the excess of quasi-perpendicular shock geometries;
    \item the decreased fraction of proton-accelerating shocks potentially has an effect in reducing the $\gamma$ emission linked to proton interactions, thus possibly explaining the lack of detection by \textit{Fermi}-LAT, while leaving the acceleration of CR electrons almost unchanged (Section \ref{sec:obs}, Figure \ref{fig:gammarad3}).
\end{enumerate}
To conclude, we have found preliminary results suggesting that the role of obliquity is far from negligible in the process of particle acceleration and signature emission of CRs, and that different cosmic environments may be characterized by systematically different regimes of shock obliquity and local plasma parameters, confirming earlier findings by \citet{wi17}. 

Our modelling of CR acceleration across cosmic environments has implicitly assumed that the topological properties of magnetic fields resolved by our simulations  ($\sim 10^{21}\ \mathrm{cm}$) remain unvaried down to the scales where DSA and SDA take place ($\sim 10^{16}\ \mathrm{cm}$).
This is certainly a gross oversimplification, which we consider only as first step towards a gradual multi-scale approach for future PIC simulations, which will allow us to bridge the tremendous gap in spatial scales that separate the ``micro'' scales at which acceleration takes place, to the ``macro'' scales that radio telescopes can observe.

\section*{Acknowledgements}
We thank both our reviewers, A. Bykov and an anonymous referee, for the useful scientific feedback on our paper. The cosmological simulations were performed with the \textsc{Enzo} code (\url{http://enzo-project.org}), which is the product of a collaborative effort of scientists at many universities and national laboratories, and the 
initialization of the nested cosmological simulations were performed with the \textsc{MUSIC} code {\url{https://www-n.oca.eu/ohahn/MUSIC/}}. S.B and F.V. acknowledge financial support from the ERC  Starting Grant ``MAGCOW'', no. 714196. D.W. is funded by the Deutsche Forschungsgemeinschaft (DFG, German Research Foundation) - 441694982. The simulations on which this work is based have been produced on Piz Daint supercomputer at CSCS-ETHZ (Lugano, Switzerland) under projects s701 and s805 and on the J\"ulich Supercomputing Centre (JFZ). The authors gratefully acknowledge the Gauss Centre for Supercomputing e.V. (\url{www.gauss-centre.eu}) for supporting this project by providing computing time through the John von Neumann Institute for Computing (NIC) on the GCS Supercomputer JUWELS at J\"ulich Supercomputing Centre (JSC), under projects no. hhh42  and  stressicm (PI F.V.)) as well as hhh44 (PI D.W.). We also acknowledge the usage of online storage tools kindly provided by the INAF Astronomical Archive (IA2) initiave (\url{http://www.ia2.inaf.it}). We thank D. Soler for his useful scientific feedback, C. Gheller for reading the manuscript and providing valuable advice, M. Br\"{u}ggen for his support in the first production of the Chronos++ suite of simulations employed in this work and A. Mignone for the helpful discussion.

\subsection*{Data availability}
Relevant samples of the input simulations used in this article and of derived quantities extracted from our simulations are stored via EUDAT and can be publicly accessed through this URL: \url{https://cosmosimfrazza.myfreesites.net/scenarios-for-magnetogenesis}.

\bibliographystyle{mnras}
\bibliography{franco2}

\appendix

\section{On the conservation of B-parallel in MHD shocks}
\begin{figure}
\includegraphics[clip,trim={0.2cm 0.5cmcm 0.3cm 0cm}]{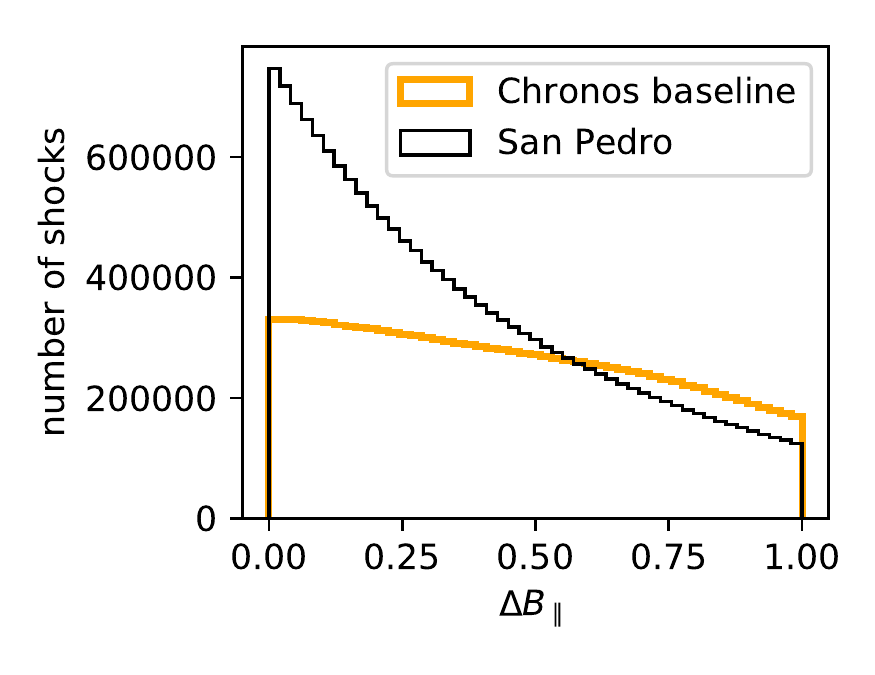}
\caption{The Figure shows the distribution of $\Delta B_{\parallel}$, i.e. how much the parallel component of magnetic fields is measured to change from pre-shock to post-shock cell for each shock in the Chronos baseline run and in the San Pedro clusters.}
\label{fig:errB}
\end{figure}

According to the standard shock jump conditions in ideal MHD  \citep[see][]{fitzpatrick2014plasma} the regions crossed by shocks experience a modification in the strength of the magnetic field perpendicular to the shock normal ($B_{\perp}$) due to compression, while the parallel component ($B_{\parallel}$) is conserved. 
Therefore, in principle the conservation of $B_{\parallel}$ going from the up-stream to the down-stream region can be considered as a way to constrain the identification of the propagation direction of numerical shocks. However, we tested that in practice this operation is prone to typically large numerical errors, due to the fact that multiple flows can affect the evolution of cell values over one timesteps, and that in the most general case the change in $B_{\parallel}$ and $B_{\perp}$ is not uniquely due to shocks.
We estimated the variation in the parallel component of the magnetic field as

\begin{equation}
\mathrm{\Delta B_{\parallel}}=\left|\frac{B_{\parallel \mathrm{post}}-B_{\parallel \mathrm{pre}}}{B_{\parallel \mathrm{pre}}}\right|.
\end{equation}
In Figure \ref{fig:errB}, we show the distribution of $\Delta B_{\parallel}$ for the two sets of simulations. 
The lack of conservation in $B_{\parallel}$ is linked to the resolution of the simulation, since the smaller the cell size the smaller the time interval that separates the pre-shock and post-shock cells. Assuming a typical velocity of $\sim 10^2-10^3\ \mathrm{km\ s^{-1}}$ for accreting matter, for the Chronos simulations (resolution of $\approx 83\ \mathrm{kpc}$) the timescale is of the order of $10^8-10^9\ \mathrm{yr}$; for the higher-resolution simulations ($\approx 25\ \mathrm{kpc}$) the timescale is $\sim 5\cdot 10^7-5\cdot 10^8\ \mathrm{yr}$. As a consequence, the higher accuracy of the pre-shock cell identification in San Pedro clusters leads to a better conservation of $B_{\parallel}$, while the values of obliquity obtained with Chronos may not be fully reliable. Nonetheless, we already established that the two sets of simulations are mostly in agreement in terms of obliquity (e.g. Figure \ref{fig:parpersp}), so we consider our estimates for Chronos to be valid to a first approximation.

\label{app:parall}

\section{On the distinction between perpendicular and parallel shocks}
\label{app:parper}
\begin{figure}
\includegraphics[clip,trim={0.3cm 0.5cm 0.3cm 0}]{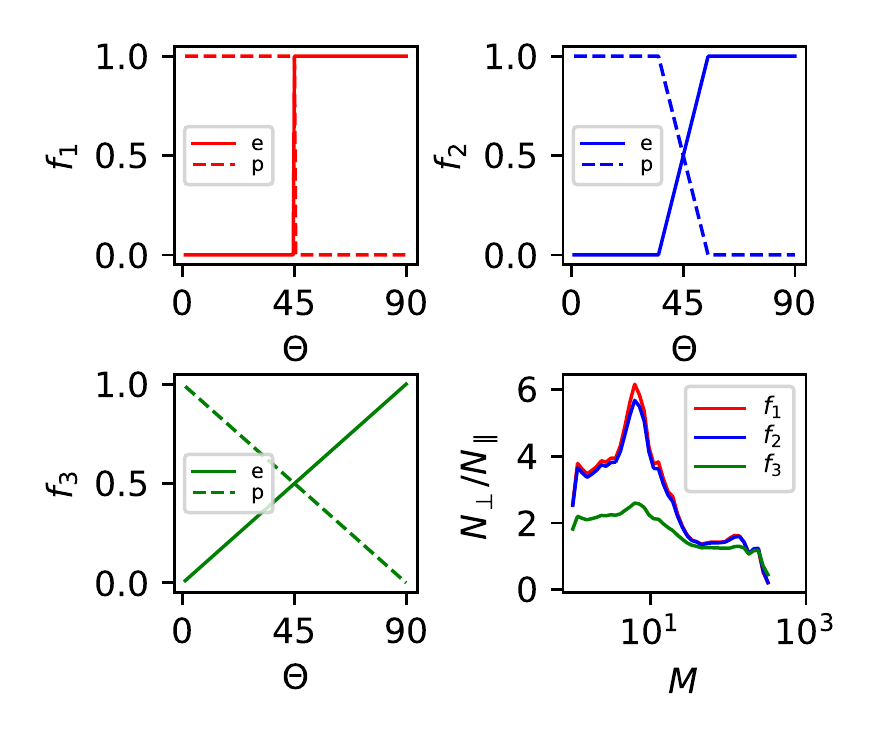}
\caption{Plots of the functions $f_1$, $f_2$ and $f_3$ determining the fraction of electrons (solid lines) and protons (dashed lines) of the total amount of CR that a shock with a certain obliquity can accelerate. The lower-right panel shows the effect of the different functions $f$ on the ratio ${N_{\perp}}/{N_{\parallel}}$ for shocks in the San Pedro clusters.}
\label{fig:f}
\end{figure}

In this analysis we identified as quasi-perpendicular shocks the ones having $\Theta>45^{\circ}$, while quasi-parallel shocks are the ones with $\Theta <45^{\circ}$. PIC simulations \citep{guo14,2014ApJ...783...91C} showed that this transition should be a smoother function of $\Theta$. We show how the ratio of number of perpendicular over parallel shocks in the San Pedro runs varies for three different functions $f$ ranging from $0$ to $1$ that parametrize the ability of a shock to accelerate electrons and protons. $f_1$ is the step function adopted for the analysis; $f_2$ assumes a linear transition from $0$ to $1$ in the interval $35^{\circ}<\Theta<55^{\circ}$; $f_3$ assumes a linear transition from $0$ to $1$ in the interval $0^{\circ}<\Theta<90^{\circ}$. Figure \ref{fig:f} shows that there is no significant discrepancy in the trend of the ratio as a function of Mach number from $f_1$ to $f_2$, while $f_3$ would introduce a non-negligible change. However, according to PIC simulations, the most accurate function should be similar to $f_2$, whose effect on ${N_{\perp}}/{N_{\parallel}}$ is approximately the same as $f_1$'s. 

\section{Spurious small-scale waves}
\label{app:waves}
\begin{figure}
\includegraphics[width=\columnwidth,clip,trim={2cm 1cm 1cm 1cm}]{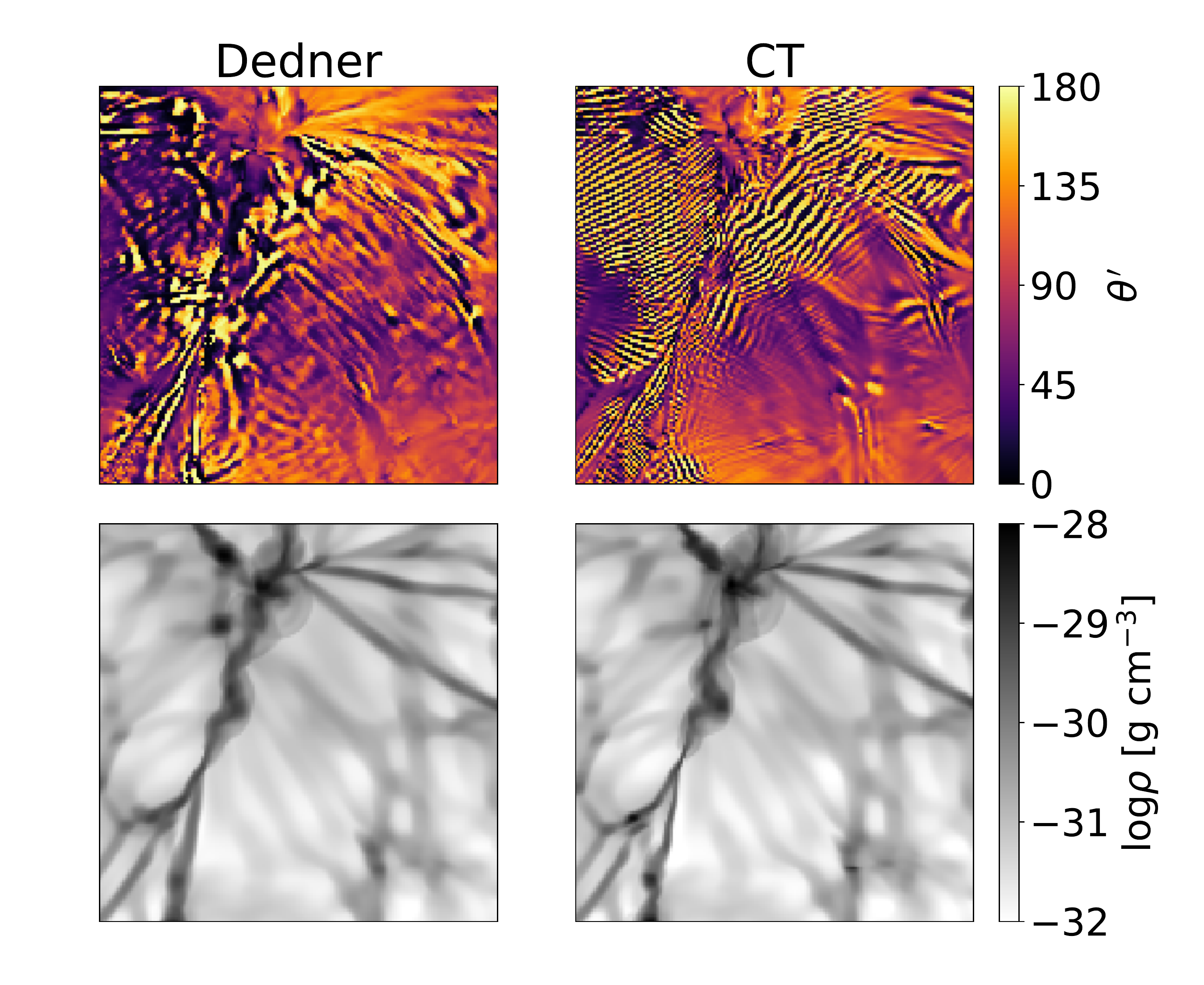}
\caption{Angle formed by velocity and magnetic field (``pseudo-obliquity'') in a slice of volume for simulations run with Dedner cleaning and CT in the top panels. Density slice of the corresponding simulations in the bottom panels.}
\label{fig:pse}
\end{figure}

\begin{figure}
\includegraphics[clip,trim={0.3cm 0.5cmcm 0.3cm 0.2cm}]{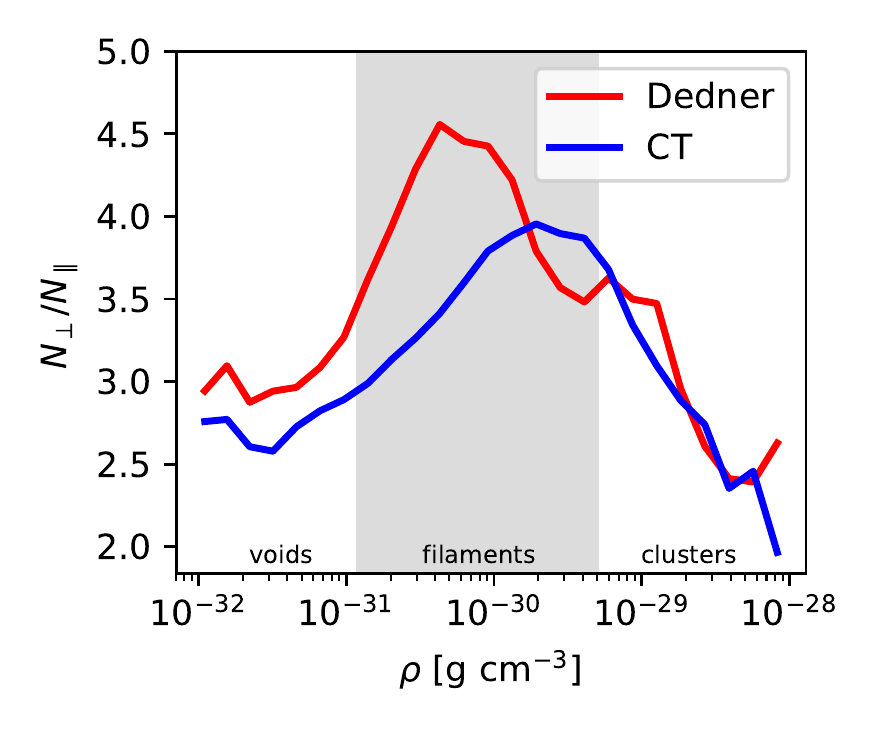}
\caption{Trend of $N_{\perp}/N_{\parallel}$ as a function of gas density for the $400^3$ simulations, run with Dedner cleaning (in red) and CT (in blue). The $\rho$-axis is divided into density intervals identifying voids, filaments and clusters.}
\label{fig:dedct}
\end{figure}

We investigated a quantity which is strictly related to obliquity, but can be defined for each cell, not only for shocked ones: this ``pseudo-obliquity'' $\theta'$ is the angle between the gas velocity and the magnetic field. We found that this quantity behaves peculiarly, i.e. ``stripes'' of null and straight angles are formed in rarefied regions of the four Chronos runs, similar to waves crossing the ICM. When $\theta'\approx 0^{\circ}$ or $\theta'\approx 180^{\circ}$, $\mathbfit{v}$ and $\mathbfit{B}$ are aligned and they assume alternately either concordant or inverted direction: by analysing the orientation of velocity and magnetic field vectors we have determined that this phenomenon is associated to the inversion of the magnetic field. We suspected that these features could be linked to the divergence cleaning method \citep{ded02}, so we analyzed two smaller simulations ($400^3$ cells, with a $147\ \mathrm{kpc}$ resolution and initial conditions similar to the baseline from Chronos), run respectively with Dedner cleaning and constrained transport\footnote{We remark that in the CT run we encountered some numerical issues of unclear origin in the magnetic field computation by \textsc{Enzo} in high-density regions, which	discouraged us from using this solver for the Chronos runs: however, here we want to compare the behavior of the magnetic field in rarefied environments, which appears to be reliable.} (CT) \citep{2010ApJS..186..308C} and otherwise identical. In Figure \ref{fig:pse}, we show the values of $\theta'$ in a slice of the volume for the two simulations, along with the corresponding density: analogous features are present in both cases, but the morphology of the stripes is quite different. In order to ascertain that the solver does not drastically affect the obliquity estimation, we computed $\theta$ for shocks in both cases and verified that the relative abundance of perpendicular and parallel shocks is overall compatible in most environments, as can be seen in Figure \ref{fig:dedct}, where the trend of $N_{\perp}/N_{\parallel}$ is shown as function of density.
Although we do not have a physical or numerical explanation for this phenomenon, this may have repercussion only on low-density regions, which host a very small fraction of the totality of shocks. Given these considerations, the previous analysis should only marginally be affected by this issue: precisely, the obliquity of shocks in rarefied regions (with very high Mach numbers) is not completely reliable. 

\section{Comparison with interstellar medium shocks}
\label{app:xi}
In this Section, we illustrate the work on simulations of the ISM performed by \citet[][]{Soler2017} and show how some of the considerations they made applied on our work as well. As anticipated in Section \ref{sec:res_sp}, we found that the magnetic field tends to arrange around low-density structures in a direction which is parallel to the structure itself, due to the action of shear motions. \citet[][]{Soler2017} encountered this phenomenon while simulating turbulent molecular clouds: they considered the angle $\phi$ formed by the density gradient $\nabla\rho$ and the magnetic field $\mathbfit{B}$ and defined the quantity $\xi$ (relative orientation parameter) as the difference between the number of cells where $\left|\cos\theta\right|<0.125$ and the number of cells where $\left|\cos\theta\right|>0.875$: thus $\xi>0$ where $\nabla\rho$ is quasi-perpendicular to $\mathbfit{B}$ and $\xi<0$ where they are quasi-parallel\footnote{We point out that while for random vectors in space the probability distribution of $0\leq\phi\leq \pi$ is $\propto\sin\phi$, all values assumed by $-1\leq\cos\phi\leq 1$ are equiprobable, so a random distribution returns $\xi=0$.}. 
\begin{figure}
\includegraphics[clip,trim={0.3cm 0.5cm 0.3cm 0.2cm}]{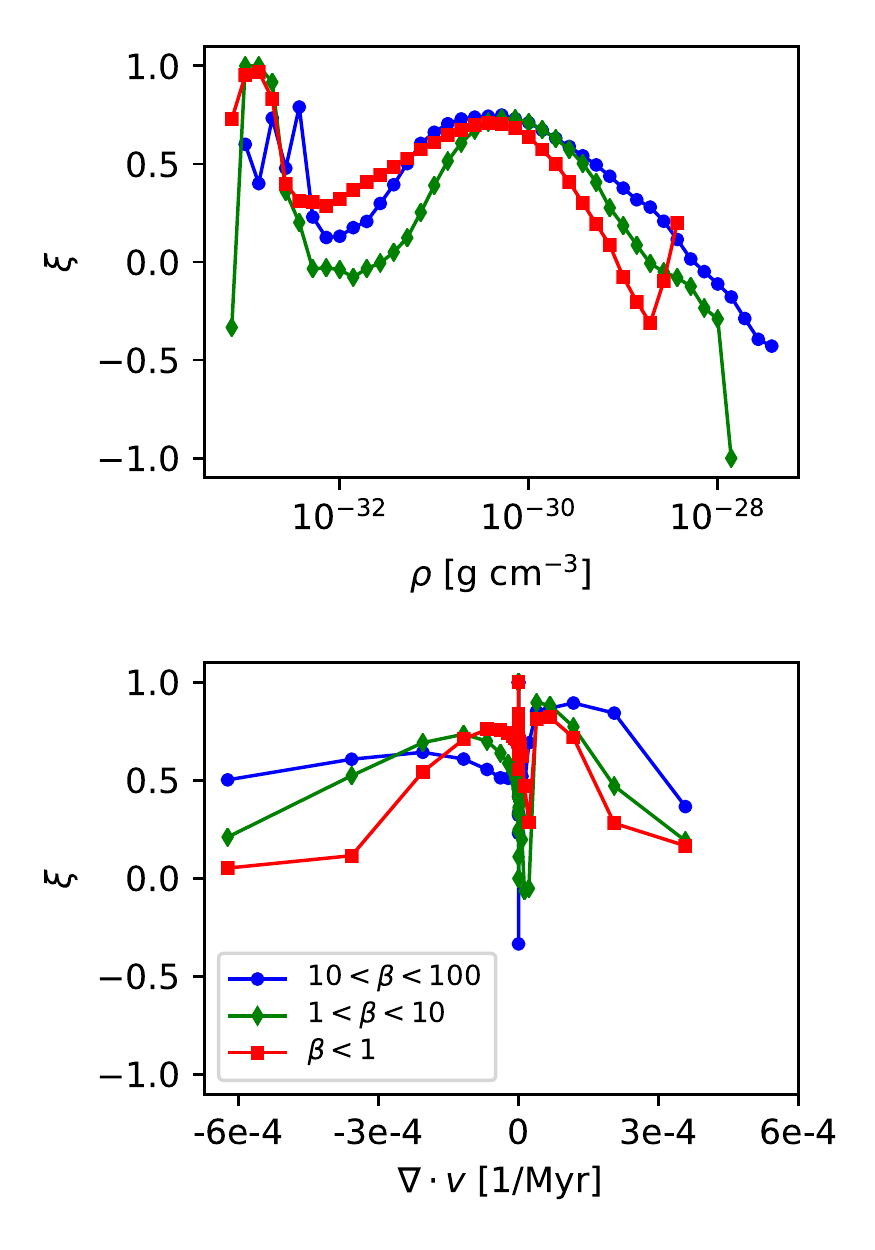}
\caption{Comparison to Figure 1 and 2 in \citet[][]{Soler2017}: relative orientation parameter $\xi$ as a function of density and velocity divergence for the Chronos baseline simulation for different $\beta$ intervals.}
\label{fig:xi}
\end{figure}
In Figure \ref{fig:xi}, we replicate the plots from Figure 1 and 2 in \citet[][]{Soler2017} of the relative orientation parameter as a function of the gas density, which identifies the different structures, and as a function of the velocity divergence, which is strictly related to the shock strength. The behavior of $\xi$ is evaluated at varying values of the plasma $\beta$. Even though they considered a very different range of densities and velocities, we found similar results in the Chronos baseline simulation. Except for very rarefied regions, possibly affected by numerics, we note the same behavior for $\xi$, which is positive in most of the volume, i.e. there are more perpendicular configurations with respect to parallel ones, with a decreasing trend towards higher densities, towards more negative values of velocity divergence and towards lower $\beta$ (i.e. more magnetized plasmas).

\bsp	%
\label{lastpage}

\end{document}